\def\rn{}
\def\nn#1 #2{#2. #1}				% Name with 1 initial
\def\nnn#1 #2 #3{#2. #3. #1}			% Name with 2 initials
\def\nnnn#1 #2 #3 #4{#2. #3. #4 #1}		% Name with 3 initials
\def\nnnnn#1 #2 #3 #4 #5{#2. #3. #4 #5. #1}	% Name with 4 initials
\def\dualand{ and\hbox{ }}				
\def\multiand{, and\hbox{ }}				
\def\rf#1;#2;#3;#4;#5 {{\frenchspacing\par\rn#1, #3 {\bf #4}, #5 (#2). \par}}
\def\rg#1;#2;#3;#4;#5;#6 {{\frenchspacing\par\rn#1, #3 {\bf #4}, #5 (#2). \par}}
\def\rfproc#1;#2;#3;#4;#5;#6 {{\frenchspacing\par\rn#1 #2, in {\it #3}, ed. #4 (#5: #6)\par}}
\def\rfprocp#1;#2;#3;#4;#5;#6;#7 {{\frenchspacing\par\rn#1 #2, in {\it #3}, ed. #4 (#5: #6), p#7\par}}
\def\rg#1;#2;#3;#4;#5;#6 {\par\rn#1 #2, {\it #3}, {\bf #4}, #5 (``#6'') \par}
\def\rf#1;#2;#3;#4;#5 {\par\rn#1, {\it #3}, {\bf #4}, #5 (#2)\par}
\def\rfbook#1;#2;#3;#4;#5 {{\frenchspacing\par\rn#1, {\it #3} (#4: #5, #2)\par}}
\def\rfproc#1;#2;#3;#4;#5;#6 {{\frenchspacing\par\rn#1 #2, in {\it #3}, ed. #4 (#5: #6)\par}}
\def\rfprocp#1;#2;#3;#4;#5;#6;#7 {{\frenchspacing\par\rn#1 #2, in {\it #3}, ed. #4 (#5: #6), p#7\par}}
\def\rfprep#1;#2;#3 {{\par\frenchspacing\rn#1, #3 (#2)\par}}
\def\rfprepp#1;#2;#3 {{\par\rn#1 #2, #3\par}}

\def\psibra{\langle\psi|}
\def\psiket{|\psi\rangle}

\def\up{|\!\!\uparrow\rangle}
\def\down{|\!\!\downarrow\rangle}

\def\upbra{\langle \uparrow\!\!|}
\def\downbra{\langle \downarrow\!\!|}

\def\aready{|a_r\rangle}
\def\aup{|a_\uparrow\rangle}
\def\adown{|a_\downarrow\rangle}

\def\iready{|i_r\rangle}
\def\iup{|i_\uparrow\rangle}
\def\idown{|i_\downarrow\rangle}

\def\ireadybra{\langle i_r|}

\def\confusionoperator{\widehat{\hskip-0.5mm\raisebox{2pt}{\textbigcircle}\hskip-3.4mm\raisebox{0pt}{${\ddot\sim}$}}}
\def\happyoperator{\widehat{\hskip-0.5mm\raisebox{2pt}{\textbigcircle}\hskip-3.85mm\raisebox{0pt}{${\ddot\smile}$}}}

\def\tensormult{\otimes} % For now
\def\freqoperator{\widehat{F}}
\def\freqvecoperator{\widehat{\bf F}}
\def\Ihat{\widehat{I}}
\def\I{{\bf I}}
\def\C{{\bf C}}
\def\f{{\bf f}}
\def\p{{\bf p}}
\def\erf{{\mathrm{erf}}}
\def\erfc{{\mathrm{erfc}}}

\def\etal{{\frenchspacing\it et al.}}
\def\ie{{\frenchspacing\it i.e.}}
\def\eg{{\frenchspacing\it e.g.}}

\def\beq#1{\begin{equation}\label{#1}}
\def\eeq{\end{equation}}
\def\beqa#1{\begin{eqnarray}\label{#1}}
\def\eeqa{\end{eqnarray}}
\def\eq#1{equation~(\ref{#1})}
\def\Eq#1{Equation~(\ref{#1})}

\def\fig#1{Figure~\ref{#1}}

\def\Sec#1{Section~\ref{#1}}
\def\Sec#1{Section~\ref{#1}}

\def\spose#1{\hbox to 0pt{#1\hss}}
\def\simlt{\mathrel{\spose{\lower 3pt\hbox{$\mathchar"218$}}
     \raise 2.0pt\hbox{$\mathchar"13C$}}}
\def\simgt{\mathrel{\spose{\lower 3pt\hbox{$\mathchar"218$}}
     \raise 2.0pt\hbox{$\mathchar"13E$}}}
\def\simpropto{\mathrel{\spose{\lower 3pt\hbox{$\mathchar"218$}}
     \raise 2.0pt\hbox{$\propto$}}}

\def\beq#1{\begin{equation}\label{#1}}
\def\eeq{\end{equation}}
\def\beqa#1{\begin{eqnarray}\label{#1}}
\def\eeqa{\end{eqnarray}}
\def\eq#1{equation~(\ref{#1})}
\def\Eq#1{Equation~(\ref{#1})}

\documentclass[twocolumn,amsmath,nofootinbib,showpacs]{revtex4} % For astro-ph
\usepackage{amsfonts,amsbsy,textcomp,epsf} % hypernat doesn't work on my laptop
 \usepackage{graphicx}
\usepackage{epstopdf}
\begin{document}
% Include Rokicki's epsf.sty file for Encapsulated PostScript graphics
%\input{epsf.sty}

%\date{August 31, 2010}
\date{\today. To be submitted to Phys.~Rev.~D}
%\date{Submitted to Phys.~Rev.~D. June 2 2008, revised March 27 2009, accepted March 31 2009}

\title{Born in an Infinite Universe: a Cosmological Interpretation of Quantum Mechanics} 

\author{Anthony Aguirre}
\address{Dept.~of Physics \& SCIPP, University of California at Santa Cruz, Santa Cruz, CA 95064; aguirre@scipp.ucsc.edu}
\author{Max Tegmark}
\address{Dept.~of Physics \& MIT Kavli Institute, Massachusetts Institute of Technology, Cambridge, MA 02139; tegmark@mit.edu}

\begin{abstract}
We study the quantum measurement problem in the context of an infinite, statistically uniform space, as could be generated by eternal inflation.  It has recently been argued that when identical copies of a quantum measurement system exist, the standard projection operators and Born rule method for calculating probabilities must be supplemented by estimates of relative frequencies of observers. We argue that an infinite space actually renders the Born rule {\em redundant}, by physically realizing all outcomes of a quantum measurement in different regions, with relative frequencies given by the square of the wave function amplitudes.  Our formal argument hinges on properties of what we term the quantum confusion operator, which projects onto the Hilbert subspace where the Born rule fails, and we comment on its relation to the oft-discussed quantum frequency operator.  This analysis unifies the classical and quantum levels of parallel universes that have been discussed in the literature, and has implications for several issues in quantum measurement theory. Replacing the standard hypothetical ensemble of measurements repeated {\it ad infinitum} by a concrete decohered spatial collection of experiments carried out in different distant regions of space provides a natural context for a statistical interpretation of quantum mechanics.  It also shows how, even for a single measurement, probabilities may be interpreted as relative frequencies in unitary (Everettian) quantum mechanics.  We also argue that after discarding a zero-norm part of the wavefunction, the remainder consists of a superposition of indistinguishable terms, so that arguably ``collapse" of the wavefunction is irrelevant, and the ``many worlds" of Everett's interpretation are unified into one.  Finally, the analysis suggests a ``cosmological interpretation" of quantum theory in which the wave function describes the actual spatial collection of identical quantum systems, and quantum uncertainty is attributable to the observer's inability to self-locate in this collection.
\end{abstract}

% PACS, the Physics and Astronomy Classification Scheme
% http://www.aip.org/pacs/pacs03/pacs0390.html
  
\maketitle

%%%%%%%%%%%%%%%%%%%%%%%%%%%%%%%%%%%%%%%%%%%%%%
%%%%%%%%%%%%%%%%%%%%%%%%%%%%%%%%%%%%%%%%%%%%%%

\setcounter{footnote}{0}

\def\thetamin{\theta_{\rm min}}
\def\thetamax{\theta_{\rm max}}

\section{Introduction}
\label{IntroSec}

Although quantum mechanics is arguably the most successful physical theory ever invented, 
the century-old debate about how it fits into a coherent picture of the physical world shows no sign of abating.
Proposed responses to the so-called measurement problem (e.g.,~\cite{Jammer})
 include the
 Ensemble  \cite{Born26,Ballantine}, 
  Copenhagen  \cite{Bohr,Heisenberg}, 
 Instrumental \cite{Dirac,Wigner,WheelerZurek,GottfriedYan}, 
 Hydrodynamic  \cite{Madelung27}, 
 Consciousness  \cite{vonNeumann32,LondonBauer39,Wigner67,Stapp93}, 
 Bohm \cite{Bohm52},
 Quantum Logic \cite{BirkhoffNeumann36},
 Many-Worlds \cite{Everett57,EverettBook},
 Stochastic Mechanics \cite{Nelson66},
 Many-Minds \cite{Zeh70,Page},
 Consistent Histories  \cite{Griffiths84},
 Objective Collapse \cite{GRW86},
 Transactional  \cite{Cramer86},
 Modal  \cite{vanFraassen72},
 Existential  \cite{Zurek87},
 Relational  \cite{Rovelli96},
and
 Montevideo \cite{GambiniPullin09} interpretations. 
Moreover, different proponents of a particular interpretation often disagree about its detailed definition. 
Indeed, there is not even consensus on which ones should be called interpretations.

While it is tempting to relegate the measurement problem to ``philosophical" irrelevancy, the debate has both informed and been affected by other developments in physics, such as the understanding of the importance~\cite{bohm} and mechanisms~\cite{Zeh70,Zurek09,JZ85,SchlosshauerBook} of decoherence, experimental 
probes of quantum phenomena on ever larger scales, and the development of cosmology.

Putting quantum mechanics in the context of cosmology creates new issues.  How can we apply the measurement postulate at times before measuring apparatuses -- or, in some interpretations consciousness -- existed? How can we split the universe into a system and measuring apparatus if the system is the entire universe itself? And cosmology may even be intertwined with quantum measurements here-and-now.  
In an infinite universe, our {\em observable} universe (the spherical region from which light has had time to reach us during the past 14 billion years) may be just one of many similar regions, and even one of many {\em exact} copies -- and in fact many versions of the widely-accepted standard inflationary cosmology provide just such a context~\cite{GarrigaVilenkin01}, as we shall discuss.  Don Page has recently argued that pure quantum theory and the textbook Born rule cannot produce outcome probabilities for multiple replicas of a quantum experiment~\cite{Page09,Page09b,Page10}, and must therefore be augmented in a cosmological scenario in which such replicas exist.

It is therefore timely to further investigate how such a cosmology impacts the quantum measurement problem, and this is the aim of the present paper. We will argue that cosmology is not merely part of the problem, but also part of the solution. 
Page essentially showed that to calculate the probability that a measurement in a collection of $N$ identical experiments has a given outcome, one must supplement the standard Born rule for assigning probabilities to measurement outcomes, based on projection operators, by a rule that assigns probabilities to each of the events ``the experiment was performed by the $k$th observer, $k = 1...N$".
We will go further, and argue that if one identifies probabilities with the relative frequencies of experimental outcomes in three-dimensional space,  the measurement postulate (Born's rule) becomes superfluous, as it emerges directly from the quantum Hilbert space formalism and the equivalence of all members of an infinite collection of exact replicas.
We will see that this is intimately linked to classic frequency operator results \cite{Finkelstein,Hartle68,HartleSrednicki,Farhi89,Gutmann95}, except that a fictitious infinite sequence of identical measurements is replaced by an actually 
existing spatial collection.  It is also closely related to arguments\footnote{In brief, these works argue that a complete description of the universe does not single out a particular place.  So instead of describing what happens ``here" it describes an ensemble (in Gibbs' sense) of identical experiments uniformly scattered throughout an infinite (expanding) space (a cosmological ensemble).  It follows from a description of the measuring process in which the measuring apparatus is assigned a definite macrostate but not a definite microstate that the measurement outcomes and their relative frequencies in the cosmological ensemble coincide with those given by the measurement postulate. (A similar idea was independently put forward in schematic form by~\cite{vilenkinreplicas}.)} by~\cite{LayzerBook,Layzer}.

This potential cosmological connection between probabilities in quantum mechanics and the relative frequencies of actual observers is relevant to most of the above-mentioned quantum interpretations.
It is particularly interesting for Everett's Many Worlds Interpretation (MWI), where we will argue that it eliminates the perplexing feature that, loosely speaking, some observers are more equal than others. To wit, suppose a spin measurement that should yield ``up'' with probability $p=0.5$ is repeated $N=10$ times. According to the MWI, the final wavefunction has $2^N=1024$ terms, each corresponding to
an equally real observer, most of whom have measured a random-looking sequence of ups and downs. This suggests that quantum probabilities can be given a simple frequentist interpretation.
However, for an unequal probability case such as $p=0.001$, the final wavefunction still has $2^N$ terms corresponding to
real observers, but now most of them have measured approximately 50\% spin up and concluded that the Born rule is incorrect. (We are supposed to believe that everything is still somehow consistent because 
the observers with a smaller wave function amplitude are somehow 
``less real".) We will show that in an infinite inflationary space, probabilities can be given a frequentist interpretation even in this case.

The rest of this paper is organized as follows.
In \Sec{CosmoSec}, we describe the cosmological context in which quantum mechanics has found itself.
We then investigate how this yields a forced marriage between quantum probabilities and relative frequencies, in both a finite space (\Sec{FiniteSec})
and an infinite space (\Sec{InfiniteSec}).
Rather than launching into an intimidating mathematical formalism for handling the most general case, we 
begin with a very simple explicit example, then return to the rather unilluminating issue of how to generalize the result in Appendix~\ref{GeneralizationSec}.
In \Sec{DecohereSec} we describe how to formally describe measurement in this context, then
discuss possible interpretation of our mathematical results in \Sec{InterpretationSec}.  We discuss some open issues in \Sec{DiscSec}, and summarize our conclusions in \Sec{ConcSec}.

\section{The cosmological context}
\label{CosmoSec}

When first applying General Relativity to our universe, Einstein assumed the Cosmological Principle (CP): our universe admits a description in which its large-scale properties do not select a preferred position or direction.  This principle has served cosmology well, supplying the basis for the open, flat, and closed universe metrics that underly the highly successful Friedmann-Lema\^itre-Robertson-Walker (FLRW) Big-Bang cosmology.
  We shall argue that this principle and the interpretation of QM may be closely intertwined, with the theory of cosmological inflation as a central player.  In particular, we will discuss how {\em eternal} inflation naturally leads to a universe obeying a strong version of the CP~\cite{Layzer1,LayzerBook}, in which space is infinite and has {\em statistically uniform}\footnote{The matter distribution in space is customarily modeled as evolved from some random initial conditions, so properties in any given region must be described statistically, e.g., by the mean density, 2-point and higher correlation functions, etc. By {\em statistically uniform}, we mean that all such statistics are translationally invariant. (For example, the homogeneous and isotropic Gaussian random fields generated by inflation  -- and anything evolved from such initial conditions -- are statistically uniform in this sense.)  Another way to look at this is that the probability distribution for different realizations in a given region of space is the same as the distribution across different spatial regions in a given realization. (See~\cite{Layzer} for an extended discussion of this point and its implications.)
} 
properties.
In this context, any given finite region is replicated throughout the infinite space, which in turn requires a re-appraisal of quantum probabilities.

\subsection{The Cosmological Principle and infinite spaces}
\label{sec-cpinfinite}
	
In a finite space, the CP has a curious status: with a single realization of a finite space, there is no meaningful way for the statistical properties to be uniform.  There would, for example, always be a unique point of highest density.  We could compare our realization to a hypothetical ensemble of universes generated assuming a set of uniform statistical properties, but we could never {\em recover} these putative statistical properties beyond a certain degree of precision.  In this sense the CP in a finite space is really nothing more than an assumption (as by Einstein) that space and its contents are ``more or less" homogeneous on large scales; a precise description would require the specification an enormous amount of information. 

An infinite space is quite different: by examining arbitrarily large scales, its statistical properties can in principle be assessed to arbitrarily high accuracy about any point, so there is a precise sense in which the properties can be uniform.  Moreover, if (as the holographic principle suggests) a region of some finite size and energy can only take on a fixed finite set of possible configurations, then the full specification of a statistically-uniform infinite space would require {\em only} those statistics.  This implies~\cite{Layzer1,LayzerBook} that in contrast to a finite system, there would be only one possible realization of such a system, as any two systems with the same statistical properties would be indistinguishable.

The CP might be taken as postulated symmetry properties of space and its contents, consistent with the near uniformity of our observed universe.  In an infinite (open or flat) FLRW universe, these symmetries can be {\em exact} in the above sense, and such a postulated cosmology would support the arguments of this paper beginning in \Sec{QMLinkSec}, or those of~\cite{Layzer}. 

Alternatively, we might search for some physical {\em explanation} for the near-uniformity of our observable universe. This was a prime motivation for cosmological inflation.  Yet inflation can do far more than create a large uniform region: in generic models inflation does, in fact, create an {\em infinite} uniform space.
	
\subsection{Infinite spaces produced by eternal inflation}

Inflation was devised (\cite{guth}; see~\cite{LindeRev} for some history) as a way to grow a finite-size region into an extremely large one with nearly uniform properties, and if inflation is realized in some region, it does this effectively: the exponential expansion that inflates the volume also dilutes or stretches into near homogeneity any particles or fields within the original region.  The post-inflationary properties are then primarily determined not by cosmic initial conditions, but by the {\em dynamics} of inflation, which are uniform across the region; although particular initial conditions are required for such inflation to arise, once it does, information about the initial conditions is largely inflated away.

It was soon discovered, however, that in generic models, inflation is {\em eternal}: although inflation eventually ends with probability unity at any given location, the exponential expansion ensures that the total inflating volume always increases exponentially (see~\cite{LindeRev,GuthRev,AguirreRev} for recent reviews.)  In many cases, one may think of this as a competition between the exponential expansion $\exp(Ht)$, and the ``decay" from inflation to non-inflation with characteristic time $t_{\rm decay}$.  This means that an initial inflating volume $V$ has, at some later time, inflating volume 
$\sim V\exp(3Ht)\exp(-t/t_{\rm decay}) = V\exp[(3H-t_{\rm decay}^{-1})t]$; for inflation to work at all requires the expansion to win 
for a number of $e$-foldings, implying a positive exponent; but in this case expansion will tend to win forever.\footnote{
  This is not to say that {\em every} inflation model has eternal behavior: it is not hard to devise non-eternal versions; but the need to do so deliberately in most cases suggests that eternal behavior is more generic. (An exception is hybrid inflation, which is generically non-eternal~\cite{Lindehybridnoteternal}; such models however tend to predict a scalar spectral index $n > 1$~\cite{wands}, which is in some conflict with current constraints~\cite{wmap7}.) In scenarios where inflation might take place in parallel in different parts of a complicated potential energy ``landscape'', regions of the landscape with eternal inflation will naturally outcompete those with non-eternal inflation, predicting by almost any measure that the region of space we inhabit was generated by eternal inflation.  On the other hand, it has been argued that inflation eternal inflation (\cite{bdg,arkaniswamp}) and perhaps even inflation (e.g.~\cite{tegmarkkachru}) may be difficult to realize in a landscape that is generated as a low-energy effective potential from a true high-energy quantum gravity theory.
}
The result is that eternal inflation does provide post-inflationary regions with the requisite properties, but as part of an ultimately infinite spacetime. 

It might seem that a given post-inflationary region is necessarily finite, because no matter how long inflation goes on, it can only expand a given finite initial region into a much larger yet still finite space.  
But this is not the case. General Relativity forbids any fundamental choice of time variable, but there is a physically preferred choice, which is to equate equal-time surfaces with surfaces of constant
inflaton field value (and hence constant energy density), so that the end of inflation occurs at a single time.  In eternal inflation, this choice leads to multiple disconnected surfaces on which inflation ends, {\em each one} generally being both infinite and statistically uniform.  Likewise, in each region and in these coordinates, the ensuing cosmic evolution occurs homogeneously.\footnote{Moreover, a given point on the spatial surface at which inflation ends will occur an enormously or infinitely long duration after any putative initial conditions for inflation.  Thus, insofar as inflation makes these initial conditions irrelevant, they are arguably {\em completely} irrelevant in eternal inflation.}

This occurs in all three basic types of eternal inflation:  ``open" inflation (involving quantum tunneling, and driven by an inflaton potential with multiple minima), in ``topological" inflation (driven be an inflaton field stuck around a maximum in its potential), 
and ``stochastic" inflation (in which upward quantum fluctuations of the field can overwhelm the classical evolution of the field toward smaller potential values).  These three particular scenarios are discussed in more detail in Appendix~\ref{InflationDetailsSec}, where we also provide heuristic arguments as to why infinite, statistically uniform spaces are a generic product of eternal inflation, by its very nature.\footnote{In a cosmology with a fundamental positive cosmological constant, this issue becomes more subtle, as some arguments suggest such a cosmology should be considered as having a finite total number of degrees of freedom (see, e.g.~\cite{banksfinite}).  How this can be understood consistently with the semi-classical spacetime structure of eternal inflation is an open issue.}

Thus eternal inflation, if it occurs, provides a causal mechanism for creating a space (or set of spaces) obeying a form of the CP, in the sense that each space is infinite and has uniform properties determined on average by the classical evolution of the inflaton, with statistical variations provided by the quantum fluctuations of the field during inflation.  
	
\subsection{Infinite statistically uniform space, and probabilities, from inflation}
\label{ErgodicSec}
	
	The fact that post-inflationary spacetime is infinite in eternal inflation leads to some rather vexing problems, including the ``measure problem" of how to count relative numbers of objects so that statistical predictions for the cosmic properties surrounding those objects can be made (see, e.g.,~\cite{winitskireview} for a recent review.)  This paper is not an attempt to solve that problem.  In particular, we do not address the comparison of observer numbers across regions with a different inflationary history and hence different gross properties. Rather, we will ask about what happens when we apply the formalism of quantum theory to a system in the context of a single infinite space with uniform and randomly-determined statistical properties.\footnote{While it is our assumption for present purposes, it is not a given that these questions are inseparable, as some ``global" measures would also ``induce" a measure over the otherwise uniform sub-spaces we are considering.}  
	
One of the greatest successes of cosmological inflation is that small-scale quantum fluctuations required by the Heisenberg uncertainty relation get stretched with the expanding space, then amplified via gravitational instability into cosmological large-scale structure
just like that we observe in, \eg, the galaxy distribution and in the cosmic microwave background \cite{sdsspars,wmap7}.
In a given finite cosmic region, this process creates pattern of density fluctuations representing a single realization of a statistical process with a probability distribution governed by the dynamics of inflation and the behavior of quantum fields within the inflating space.  

Eternal inflation also creates infinitely many other nearly-homogeneous regions with density fluctuations drawn from the same distribution (because the dynamics are just the same) that evolve independently of each other (because the regions are outside of causal contact if they are sufficiently widely separated, where ``widely'' means being farther apart than the horizon scale during inflation, say $10^{-24}$ m).  The resulting space then has statistically uniform properties, and the probability distribution governing the fluctuations in any single region is recapitulated as the relative frequencies of these fluctuation patterns across the actually-existing spatial collection of regions. 

Now, these inflationary fluctuations constitute the classical cosmological ``initial" conditions that determine the large-scale variation of material density and thus, e.g., the distribution of
galaxies.  Smaller-scale details of the current matter distribution (such as what you ate for breakfast) were determined 
by these same inflationary initial conditions, augmented by subsequent quantum fluctuations amplified by chaotic dynamics, etc. Because such small-scale processes (and any microscopic ``initial" conditions connected with them) are decoupled from the super-horizon large-scale dynamics giving rise to the infinite space, the overall space
should again be statistically uniform, here in the sense that the probability distribution of microstates in each finite region depends {\em not} on its location in space, but only on its macroscopic properties, 
which are themselves drawn randomly from a region-independent statistical distribution.  

In short, inflation creates an infinite set of cosmic regions, each with ``initial conditions" and subsequently-evolving properties that are characterized (and {\em only} characterized) by a statistical distribution that is independent of the choice of region.  
	
\subsection{Quantum mechanics and replicas}
\label{QMLinkSec}
	
	 Let us now make our link to everyday quantum mechanics. For a simple example that we shall follow throughout this paper, consider a spin 1/2 particle and a Stern-Gerlach experiment for measuring the $z$-component of its spin, which has been prepared in 
the state $\psi = \alpha\down+\beta\up$. Here $\alpha$ and $\beta$ are complex numbers satisfying the usual normalization condition 
$|\alpha|^2+|\beta|^2=1$.  If we assume that a finite volume region with a roughly flat background metric has a finite set of possible microscopic configurations\footnote{Meaning a finite number of meaningfully distinct ways in which the state can be specified.  Note that although the real number coefficient $\alpha$ would seem to allow an uncountably infinite set of specifications, this is misleading: the maximum von Neumann entropy $S=-{\rm Tr}\rho \log \rho$ for our system is just $\log 2$, and at most two classical bits of information can be communicated using a single qubit. We should note, however, that while our assumption is quite standard, the precise way in which the continuous $\alpha$ would over-specify the state is a subtle question that we do not address here.} (as suggested by, e.g., the holographic principle and other ideas in quantum gravity), and that our system plus experimenter configuration evolved from one of finitely many possible sets of initial conditions drawn from the distribution governing the statistically uniform space at some early time, then it follows that this configuration must be replicated elsewhere.\footnote{This does assume some additional subtleties. For example it is argued in~\cite{LayzerBook} that ``statistical predictions do not prescribe all the properties of infinite collections. ...Any outcome that occurs a finite number of times has zero probability."  In particular, an outcome that is consistent with physical laws could in principle occur in only one observable universe.} That is, there are infinitely many places in this space where an indistinguishable experimenter has prepared the same experiment using a classically indistinguishable procedure, and therefore uses the same $\alpha$ and $\beta$ to describe the initial wavefunction of her particle. The rather conservative estimate in \cite{multiverse} suggests that the nearest indistinguishable copy of our entire observable universe (``Hubble volume") is no more than $10^{10^{115}}$ meters away, and the nearest subjectively indistinguishable experimenter is likely to be much closer.\footnote{The frequency of such repetitions depends on very poorly understood questions such as the probability for certain types of life to evolve, etc.} 
We will now argue that the quantum description of this infinite set of systems sheds light on the origin of probabilities in quantum mechanics.

\section{Probabilities for measurement outcomes in a finite region}
\label{FiniteSec}

\subsection{The problem}

In a statistically uniform space, consider a finite region that is large enough to contain $N$ identical copies of our Stern-Gerlach experiment prepared in the simple above-mentioned state.

The state of this combined $N$-particle system is simply a tensor product with $N$ terms. For example, $N=3$ gives the state
\beqa{PsiEq}
\psiket&=&(\alpha\down+\beta\up) \tensormult (\alpha\down+\beta\up) \tensormult (\alpha\down+\beta\up)=\nonumber\\
       &=&\alpha^3|\!\!\downarrow\downarrow\downarrow\rangle 
        + \alpha^2\beta|\!\!\downarrow\downarrow\uparrow\rangle + ...
        + \beta^3|\!\!\uparrow\uparrow\uparrow\rangle.
\eeqa

If we order the $2^N$ basis vectors of this $2^N$-dimensional Hilbert space
by increasing number of ``up'' vectors, the vector of 
wavefunction coefficients takes the simple form 
\beq{PsiEq2}
\left(
\begin{tabular}{c}
$\langle \downarrow\downarrow\downarrow|\psi\rangle$\\
$\langle \downarrow\downarrow\uparrow  |\psi\rangle$\\
$\langle \downarrow\uparrow  \downarrow|\psi\rangle$\\
$\langle \uparrow  \downarrow\downarrow|\psi\rangle$\\
$\langle \uparrow  \uparrow  \uparrow  |\psi\rangle$\\
$\langle \uparrow  \uparrow  \downarrow|\psi\rangle$\\
$\langle \uparrow  \downarrow\uparrow  |\psi\rangle$\\
$\langle \uparrow  \uparrow  \uparrow  |\psi\rangle$
\end{tabular}
\right)
= 
\left(
\begin{tabular}{c}
$\alpha^3$\\
$\alpha^2\beta$\\
$\alpha^2\beta$\\
$\alpha^2\beta$\\
$\alpha\beta^2$\\
$\alpha\beta^2$\\
$\alpha\beta^2$\\
$\beta^3$\\
\end{tabular}
\right)
\eeq
for our $N=3$ example.  For general $N$, there are $\left({N\atop n}\right)$ terms with $n$ spins up, 
each with coefficient $\alpha^{N-n}\beta^n$.

\subsection{Probabilities in the finite region}

Suppose we would like to ask the core question: ``given that I have prepared the quantum system as described, what is the probability that I will measure $\uparrow$?"  This is more subtle than it would appear, because ``I" might be part of any one of the $N$ indistinguishable experimental setups assumed.  As argued by Page~\cite{Page09,Page09b,Page10}, if one wants to consider these $N$ experiments as a single quantum subsystem of the universe, this is problematic because in this situation there is no set of projection operators that can assign outcome probabilities purely via the Born rule.\footnote{In particular, define $\hat P_{i,\uparrow}$ and $\hat P_{i,\downarrow}$ to be operators that project onto $\up$ and $\down$ for the $i$th observer (leaving the $N-1$ other components of the product vector unchanged.)  Then Page~\cite{Page10} shows that for $N=2$, there is no state-independent projection operator $\hat P_{\uparrow}$ that gives Born-rule probabilities $P(\uparrow) = \langle \hat P_{\uparrow} \rangle$ for measuring $\uparrow$ (absent information about which particular observer one is) that are a weighted combination (with positive weights) of the probabilities $P_{i,\uparrow}$ for measuring $\up$ for each given {\em known} observer $i$.}  Thus it seems that the quantum formalism by itself is insufficient, and must be in some way supplemented by additional ingredients.

While this conflicts with the idea that quantum theory alone should suffice when applied to the whole universe, we can recover the usual Born rule results for the single system in a fairly straightforward way if we augment the Born rule as applied to the product state for the $N$ systems with probabilities assigned according to relative frequencies among the $N$ systems.\footnote{This is essentially `T5' suggested in~\cite{Page09} as one possible way to restore probabilities.}  In particular, since the $2^N$ terms are orthogonal (being a basis for the tensor-product state space) we might in principle imagine measuring the whole system, and attribute a quantum probability to each term given by its squared amplitude. Yet even if just one of these terms is ``realized",\footnote{Readers preferring the Everettian perspective can accord a probability to each of these terms as branches of the wavefunction, and make a similar argument. Note, however, that it is somewhat less satisfying because if we add up the relative frequency of observers across {\em all} branches, it is by symmetry 50\% for $\uparrow$ and 50\% for $\downarrow$; this is the uncomfortable issue of some observers being more real than others noted in the introduction.}
there is still uncertainty as to which spin is measured, because there is complete symmetry between the $N$ indistinguishable measuring apparatuses.  You should thus accord a probability for $\uparrow$ given by the relative frequencies of $\uparrow$ and $\downarrow$. 

The total probability $P_\uparrow$ of measuring $\uparrow$ would then come from a combination of quantum probabilities and frequentist estimates of probability, using $P(A) = \sum_i P(B_i) P(A|B_i)$ where $P(A|B_i)$ is the conditional probability of $A$ given $B_i$. Thus
\beq{QuantClassProb}
P_\uparrow = 
\sum_{n=0}^N \left({N\atop n}\right) (\beta^*\beta)^n (\alpha^*\alpha)^{(N-n)}\frac{n}{N}.
\eeq
Each term in this sum is just the binomial coefficient $f(n; N,p)$ with $p=\beta^*\beta$
(the quantum probability from \eq{PsiEq2}
of getting $n$ $\uparrow$-factors) times $n/N$ (the probability
that among the $N$ identical observers, you are one of the $n$ who observed $\uparrow$).
Mathematically, this sum simply computes $1/N$ times the mean of the binomial distribution, which is $Np$,
giving $P_\uparrow=p$.  In this way, the standard Born rule probability $\beta^*\beta$ to measure $\uparrow$ is recovered, using a {\em combination} of the Born rule applied to the $2^N$-state superposition, and the relative frequencies of $\up$ and $\down$ contained in each (product) state in that superposition.

\subsection{Frequency and confusion operators}

If the reality of indistinguishable systems has forced us to augment quantum probabilities with probabilities based on observer frequencies, as above, it is very interesting to 
examine more carefully how these two notions of probability connect.  To do so, 
let us now define two Hermitean operators on this Hilbert space, both of which are diagonal in this basis.
The first is the {\it frequency operator} $\freqoperator$ introduced by
\cite{Finkelstein,Hartle68,Farhi89,Gutmann95}, which multiplies each basis vector by 
the fraction of the arrows in its symbol that point up; for our $N=3$ example, 
\beq{FrequencyOperatorxample}
\freqoperator=
{1\over 3}\left(
\begin{tabular}{cccccccc}
0	&0	&0	&0	&0	&0	&0	&0\\
0	&1	&0	&0	&0	&0	&0	&0\\
0	&0	&1	&0	&0	&0	&0	&0\\
0	&0	&0	&1	&0	&0	&0	&0\\
0	&0	&0	&0	&2	&0	&0	&0\\
0	&0	&0	&0	&0	&2	&0	&0\\
0	&0	&0	&0	&0	&0	&2	&0\\
0	&0	&0	&0	&0	&0	&0	&3
\end{tabular}
\right)
\eeq
The second is the {\it confusion operator} $\>\confusionoperator\>$,
which projects onto those basis vectors where the spin-up fraction
differs by more than a small predetermined value $\epsilon$ from the 
Born rule prediction $p=|\beta^2|$; for our $N=3$ example with any $\epsilon<1/3$, 
\beq{ConfusionOperatorxample}
\confusionoperator=
\left(
\begin{tabular}{cccccccc}
1	&0	&0	&0	&0	&0	&0	&0\\
0	&0	&0	&0	&0	&0	&0	&0\\
0	&0	&0	&0	&0	&0	&0	&0\\
0	&0	&0	&0	&0	&0	&0	&0\\
0	&0	&0	&0	&1	&0	&0	&0\\
0	&0	&0	&0	&0	&1	&0	&0\\
0	&0	&0	&0	&0	&0	&1	&0\\
0	&0	&0	&0	&0	&0	&0	&1
\end{tabular}
\right)
\eeq
More generally, the two operators are clearly related by 
\beq{OperatorRelationEq}
\confusionoperator = \theta\left(|\freqoperator-p|-\epsilon\right),
\eeq
where $\theta$ denotes the Heaviside step function, \ie, $\theta(x)=1$ for $x\ge 0$, vanishing otherwise.

Since both $\freqoperator$ and $\>\confusionoperator\>$ are Hermitean operators, one would conventionally interpret them as observables, with
$\freqoperator$ measuring the fraction of spins that are up, and $\>\confusionoperator\>$ measuring 1 if this fraction differs by more than $\epsilon$ 
from $p$, 0 otherwise.  

Examining the norm of the state multiplied by $(\freqoperator-p)$, we find that \cite{Hartle68}
\beqa{FproofEq}
\|(\freqoperator-p)\psiket\|^2&=&\psibra(\freqoperator-p)^2\psiket=\nonumber\\
  &=&\sum_{n=0}^N\left({N\atop n}\right)(1-p)^n p^{N-n}\left({n\over N}-p\right)^2=\nonumber\\
  &=&{p(1-p)\over N}.
\eeqa
Note that as in \eq{QuantClassProb}, $\alpha$ and $\beta$ enter only in the combinations $\alpha^*\alpha=(1-p)$ and
$\beta^*\beta=p$ (because $\freqoperator$ is diagonal), and just as \eq{QuantClassProb} is the mean of a binomial distribution (divided by $N$), here the second line is simply the variance of a binomial distribution (divided by $N^2$).  

As for the confusion operator, using \eq{PsiEq2} and \eq{ConfusionOperatorxample}
we obtain
\beqa{ConfusionProofEq}
\|\>\confusionoperator\>\psiket\|^2&=&\psibra\>\confusionoperator\>\psiket=\nonumber\\
  &=&\sum_{n=0}^N\left({N\atop n}\right)(1-p)^n p^{N-n}\theta\left(\left|{n\over N}-p\right|-\epsilon\right)\nonumber\\
  &=&\sum_{|n-Np|>N\epsilon }\left({N\atop n}\right)(1-p)^n p^{N-n}\nonumber\\
  &\le&2 e^{-2\epsilon^2 N},
\eeqa
where $\theta$ again denotes the Heaviside step function, and we have used 
Hoeffding's inequality\footnote{Specifically, apply theorem 2 of~\cite{Hoeffding}, representing the binomial distribution as the sum of $N$ independent Bernoulli distributions.}
 in the last step. Thus
$\|\>\confusionoperator\>\psiket\|$ is exponentially small if $N\gg\epsilon^{-2}$; this mathematical result will prove to be important below.

For large enough $N$, our rescaled Binomial distribution approaches a Gaussian with mean $p$ and standard deviation 
$\sqrt{p(1-p)/N}$, so 
\beq{ConfusionApproxEq}
\|\>\confusionoperator\>\psiket\|^2\approx\erfc\left[\left({N\over 2 p(1-p)}\right)^{1/2}\epsilon\right].
\eeq
($\erfc$ denotes the complementary error function, \ie, the area in the Gaussian tails).

In summary, for finite $N$, the product state of our $N$ copies represents a sum of many terms.  In some of them, 
the relative frequencies of up and down states in individual members of the spatial collection closely approximate the corresponding probabilities given by BornÕs rule.
In other terms the frequencies differ greatly from the corresponding Born-rule probabilities. However, as $N$ increases, these confusing states contribute smaller and smaller amplitude, as measured by the quickly diminishing expectation value of $\confusionoperator$.

\begin{figure}[ht]
\centerline{\includegraphics[scale=0.4]{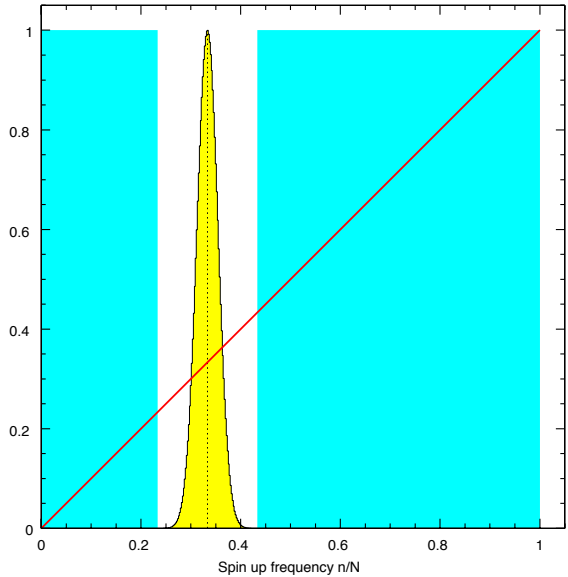}}
%\centerline{\epsfxsize=\figsize\epsffile{binomial.ps}}
\caption{\label{BinomialFig}
As a function of the spin up fraction $n/N$, the figure shows the scaled 
binomial distribution for $(n,p)=(500,1/3)$ (shaded curve) together with the
spectra of the frequency operator (diagonal line) and the 
$\epsilon=0.1$ confusion operator (shaded, equal to 1 for $|n/N-p|>\epsilon$).
}
\end{figure}

\section{Probabilites in an infinite space}
\label{InfiniteSec}

As $N \to \infty$, four key things happen. 

First, the binomial distribution governing the spread in values of the frequency operator in Eq.~\ref{FproofEq} approaches a $\delta$-function centered about $p$, so that $\|(\freqoperator-p)\psiket\|^2  = p(1-p)/N \to 0.$  

Second, the norm of the state projected by $\>\confusionoperator\>$ vanishes,
\beq{confusionvanishes}
\|\>\confusionoperator\>\psiket\|^2\to 0,
\eeq
with exponentially rapid convergence as per \eq{OperatorRelationEq}. This is illustrated in Fig.~\ref{BinomialFig}: in the \eq{ConfusionProofEq} sum,
the $\theta$-term vanishes outside the shaded area, whereas the remaining binomial distribution 
factor (the plotted curve) gets ever narrower as $N\to\infty$, eventually ending up almost entirely 
inside the region where $\theta=0$. (That $\psiket$ is an eigenvector of $\>\confusionoperator\>$ also follows directly from 
$\psiket$ being an eigenvector of $\freqoperator$ and the fact that 
$\>\confusionoperator\>$ is a function of $\freqoperator$ as per \eq{OperatorRelationEq}.
However, the figure also illustrates that 
convergence is much faster (indeed exponential) 
for the confusion operator $\>\confusionoperator\>$ 
than for the frequency operator $\freqoperator$ case, since the spectrum of the former
is constant (zero) near the limit point.)

Third, let us define the complement of the confusion operator, which projects onto the 
states where the up fraction is within $\epsilon$ of the Born rule prediction $p$:
\beq{HappyOperatorDefEq}
\happyoperator \equiv \Ihat-\confusionoperator\>,
\eeq
where $\Ihat$ is the identity operator.
Both $\confusionoperator$ and $\happyoperator$ are projection operators, and they by definition 
satisfy the relations 
${\>\confusionoperator\>}^2 = \>\confusionoperator\>$,~~
${\happyoperator\>}^2 = \>\happyoperator\>$,~ and 
$\>\confusionoperator\>\>\happyoperator\> =
\>\happyoperator\>\>\confusionoperator\> =
0$.

Now let us decompose our original state into two orthogonal components:
\beq{DecompositionEq}
\psiket = \confusionoperator\>\psiket  + (\Ihat-\confusionoperator\>)\psiket = \confusionoperator\>\psiket + \happyoperator\>\psiket.
\eeq
Substituting \eq{confusionvanishes} now implies the important result that 
\beq{BornApproxEq}
\psiket\to\happyoperator\>\psiket\quad\hbox{as}\quad N\to\infty,
\eeq
with convergence in the sense that the correction term approaches zero Hilbert space norm as $N\to\infty$.
But the right-hand-side $\>\happyoperator\>\psiket$ is a state which by definition is a superposition only of states where the relative frequencies of $\up$ and $\down$ are in precise accord with Born-rule probabilities, with an up-fraction as close to $p=|\beta|^2$ as we chose to require with our $\epsilon$-parameter. 
If we now assume that the outcome of a single quantum measurement is one of the measured observableÕs eigenvalues, almost all components of the superposition of pre-measurement states weÕve been discussing represent post-measurement states in which the relative frequencies of eigenstates are equal to the corresponding Born probabilities for the possible outcomes of a single measurement. (Measurement and decoherence are discussed in the next section.) 

Finally, as $N\to\infty$, $\psiket$ approaches (again in the sense that the correction term approaches zero Hilbert space norm) a state where every element in the grand superposition 
becomes statistically indistinguishable from the others, in the following sense.  Suppose, as above, that each term in the grand superposition is taken to represent a collection of identical apparatuses that have each registered a definite outcome corresponding to either $\up$ or $\down$.
Consider a volume of enormous but finite radius $R$ that contains some large number $M$ of our identical experiments,
order them any way you like and write down their readings
as a sequence such as $...\uparrow\uparrow\downarrow\downarrow\downarrow\uparrow\downarrow...$, 
and let $n_\downarrow$ and $n_\uparrow$ 
denote the number of readings of $\downarrow$ and $\uparrow$, respectively $(n_\downarrow=M-n_\uparrow)$.
In our infinite space, just as there are infinitely many realizations of a single experiment, there will be infinitely many 
such spherical regions, in each of which one of the $2^M$ possible outcomes is realized.  Across this collection, 
these outcomes will be realized with frequencies that are within $\epsilon$ of the
Born rule predictions $(1-p)^{M-n_\uparrow} p^{n_\uparrow}$, except for a correction term with zero Hilbert space norm.
This follows from the exact same argument given above, generalized to the case of multiple outcomes as we do below in
Appendix~\ref{GeneralizationSec}. Since we must have $(1-p)^{M-n_\uparrow} p^{n_\uparrow} > 0$ for any sequence
that actually occurs in some sphere in some branch of the wavefunction, a corollary is that this exact same 
sequence will occur in {\it every} branch of the wavefunction (after neglecting the correction term with zero Hilbert space norm), regardless of how vast this sphere is.
Moreover, the two-outcome result implies that the relative frequencies in a randomly selected volume in a given branch give 
no information whatsoever about which branch it is in, because they all have the same average frequencies. 
Finally, the multiple outcomes result tells us that 
even if we compute any finite amount of statistical data from a 
truly infinite volume (by computing what fraction of the time various combinations of outcomes occur in the infinite volume), 
the different branches remain statistically indistinguishable.  In just the sense of Sec.~\ref{sec-cpinfinite}, the different branches are equivalent to different realizations of an infinite universe with the same statistical properties, and therefore cannot be told apart.

\section{Measurement and decoherence}
\label{DecohereSec}

In our discussion above we have mentioned ``outcomes" of experiments. It is generally agreed that in a measurement process, the post-measurement state of the measurement system is encoded in the degrees of freedom of a macroscopic device (say the readout of a Stern-Gerlach apparatus, the position of a macroscopic pointer, or the brain state of an observer).  We can sketch how this process plays out in the simple example and cosmological context of this paper by considering, along with the set of replica quantum systems (each a single-spin system represented by $\alpha\up + \beta\down$), a corresponding set of indistinguishable measuring devices in a `ready' state just prior to measurement.  Following the scheme of Von Neumann \cite{VN32},  if the 
apparatus is in a particular `ready' state $\aready$, independent of the system's state, then 
interaction between the system and the apparatus causes the combined system to evolve
 into an entangled state:
\beqa{vnpremeasure}
(\alpha\up + \beta\down)\aready \longrightarrow \alpha\up \aup + \beta\down\adown,
\eeqa
where $ \aup$ and $\adown$ are states of the apparatus in which it records an `up' or `down' measurement.

Let us consider a set of $N$ perfect replicas of this setup\footnote{In appendix~\ref{RhoAppendix} we generalize this discussion to the arguably more relevant case where the measuring devices are {\em macroscopically} indistinguishable, and in particular, described by the same density matrix.} that exist in an infinite statistically uniform space for just the same reason that there are copies of  $\alpha\up + \beta\down$. Following the same reasoning as in \Sec{FiniteSec}, we can consider the product state, which looks like
\beqa{PsiEqApp}
\psiket&=&(\alpha\down+\beta\up)\aready \tensormult (\alpha\down+\beta\up)\aready \tensormult ...
\eeqa
Now, after the interaction between the system and apparatus described by \eq{vnpremeasure}, and following the exact same reasoning as  \Sec{FiniteSec} and \Sec{InfiniteSec}, when $N\to\infty$, our product state becomes an infinite superposition of terms, all of which (except for a set of total Hilbert space norm zero) look like
\beqa{PsiEqInfApp}
...\up\aup \tensormult\up\aup \tensormult\down\adown\tensormult\up\aup \tensormult\down\adown...,
\eeqa
where the relative frequencies of the terms $\up\aup$ and $\down\adown$ are given by $|\alpha|^2$ and $|\beta|^2$, respectively.  

In this way, the interaction between system and apparatus has evolved $|\psi\rangle$ from a superposition containing infinitely many identical apparatuses into one of statistically indistinguishable terms, where each term describes two different sets of apparatuses: one in which each apparatus reads `up', and one in which each reads `down', with relative frequencies $|\alpha|^2$ and $|\beta|^2$.

Now, each system described by \eq{vnpremeasure} will further interact with the degrees of freedom making up its environment, which we assume have a random character.  This causes the local superposition to undergo decoherence
\cite{Zeh70,Zurek09,SchlosshauerBook}.
This decoherence is typically quite rapid, with timescales of order $10^{-20}$ seconds being common \cite{JZ85,collapse,brain}.
Let us consider the effect of this decoherence on the density matrix $\rho$ of the full $N$-particle system.
Initially, $\rho=\psiket\psibra$, where $\psiket$ is the pure $N$-particle state given by \eq{PsiEqApp}. 
Since each apparatus is rapidly entangled with its own local environment, all degrees of freedom of the full $N$-particle 
system decohere on the same rapid timescale.
This means that when we compute the resulting $N$-particle reduced density matrix $\rho$ by partial-tracing the global density matrix over the other (environment) degrees of freedom, 
it becomes effectively diagonal in our basis
(given by Equation~\ref{PsiEq2}). The vanishing off-diagonal matrix elements are those that connect different pointer states in the grand superposition, and also therefore in the individual systems; thus any quantum interference between different states becomes unobservable.

In other words, decoherence provides
its usual two services: it makes quantum superpositions for all practical purposes 
unobservable in the ``pointer basis'' of the measurement \cite{Zurek09}, and it dynamically determines which basis this is 
(in our case, the one with basis vectors like the example in \Eq{PsiEqInfApp}).  

\section{Interpretation}
\label{InterpretationSec}

Application of quantum theory to the actually existing infinite collection of identical quantum systems that is present in an infinite statistically uniform space (such as provided by eternal inflation) leads to a very interesting quantum state.  As long as we are willing to neglect a part of the wavefunction with vanishing Hilbert-space norm, then we end up with a superposition of a huge number of different states, each describing outcomes of an infinite number of widely separated identical measurements in our infinite space. In {\em all} of them, a fraction $p=|\alpha|^2$ of the observers will have measured spin up.\footnote{For readers who are concerned whether infinity should be accepted as a meaningful quantity in physics,
it is interesting to also consider the implications of a very large but finite $N$.
In this case, the Hilbert space norm of the wavefunction component where the Born rule appears invalid is bounded by $2 e^{-2\epsilon^2 N}$,
so although it is not strictly zero, it is exponentially small as long as $N\gg \epsilon^{-2}$. For example, if $N=10^{1000}$ 
(a relatively modest number in many inflation contexts), then the Born rule probability predictions are correct to 100 decimal places 
except in a wavefunction component of norm around $10^{-10^{800}}$.
}

In this way, the quantum probabilities and frequentist observer-counting that coexisted in the finite-$N$ case have merged.
Born's rule for the relative probabilities of $\uparrow$ and $\downarrow$ emerges directly from the relative frequencies of actual observers within an unbounded spatial volume; Born's rule as applied to the grand superposition is superfluous since all give the same predictions for these relative frequencies.  In particular, the ``quantum probabilities" assumed in Eq.~\ref{QuantClassProb} to be given by $\alpha^*\alpha$ and $\beta^*\beta$ are replaced by the assumption that two vectors in Hilbert space are the same if they differ by a vector of zero norm.

One of the most contentious quantum questions is whether the wavefunction ultimately collapses or not when an observation is made.
Our result makes the answer to this question anti-climactic: insofar as the wavefunction is a means of predicting the outcome of experiments, it doesn't matter, since all the elements in the grand superposition are observationally indistinguishable.  In fact, since each term in the superposition is indistinguishable from all the others, it is unclear whether it makes sense to even call this a superposition anymore.\footnote{As well as giving identical predictions, this replacement is irrelevant because decoherence has removed any practical possibility of interference.  This does not mean that the superposition has {\em mathematically} gone away, however, any more than when decoherence is applied to a single quantum system.  Nor does it mean that mathematically the sequences are necessarily equivalent; see~\Sec{DistinguishSec}.} Since each state in the superposition has formally zero norm, there is no choice but to consider classes of them, and if we class those states with indistinguishable predictions together, then this group has total Hilbert space norm of unity, while the class of ``all other states" has total norm zero.  In term of prediction, then, the infinite superposition of states is completely indistinguishable from one quantum state (which could be taken to be any one of the terms in the superposition) with unity norm.  In this sense, a hypothetical ``collapse" of the wavefunction would be the observationally irrelevant replacement of one statevector with another functionally identical one.

 In more Everettian terms we might say that in the cosmic wavefunction, each of the ``many worlds" are the same world, where ``world'' here refers to the state of an infinite space. In the terminology of \cite{multiverse,multiverse4wheeler}, the Level I Multiverse is the same as the Level III Multiverse (and if inflation instantiates more than one solution to a more fundamental theory of physics, then the Level II Multiverse is the same as the Level III Multiverse).

All this suggests that we take a different and radically more expansive view of the statevector for a finite system: this quantum state describes not a particular system ``here", but rather the spatial collection of identically prepared systems that already exist.  This provides a real collection rather than fictitious ensemble for a statistical interpretation of quantum mechanics.  It also allows quantum mechanics to be unitary in a very satisfactory way.  Rather than the world ``splitting" into a decohered superposition of two outcomes as seen by an experimenter, infinitely many observers already exist in different parts of space, a fraction $|\alpha|^2$ of which will measure one outcome, and a fraction $|\beta|^2$ of which will measure the other. The uncertainty represented by the superposition corresponds to the uncertainty before the measurement of {\em which} of the infinitely many otherwise-identical experimenters the observer happens to be; after the measurement the observer has reduced this uncertainty.  Moreover, the ``partially real" observers in the Everett picture have vanished: observers either exist equally if they are part of the wavefunction with unity, or don't exist if they are part of the zero norm branch that has been discarded.

\section{Discussion}
\label{DiscSec}
      
The above-mentioned results raise interesting issues that deserve further work, and we comment on a couple below.

\subsection{Levels of indistinguishability}
\label{DistinguishSec}

Considering our finite or infinite sequences, in which ways are two such sequences distinguishable, and what does this mean physically?  For finite $N$, if each system is {\em labeled}, then each term in the superposition is different.  However, if we consider just {\em statistical} information such as the relative frequencies of $\up$ and $\down$, then many sequences will be statistically indistinguishable.  Moreover, if we do not label the terms, so that sequences can be reordered when they are compared, then {\em only} the relative frequencies are relevant.

Now for an infinite product state, we have argued above that once we discard a zero-norm portion, the remaining states are statistically indistinguishable using any finite amount of statistical information. If the elements are unlabeled, this statistical information is simply $p$ (the $\up$ fraction). We can ask, however, if these states are {\em mathematically} distinguishable.  To see that indeed they are, consider two such states such as $...\up\down\up\up...$, and $...\down\down\up\up...$ and represent them as binary strings with $1=\up$ and $0=\down$.  Now for each state, arbitrarily select one system, and enumerate all systems as $i=1,2,...$ by increasing spatial distance from this central element.  Each state then corresponds to a real number in binary notation such as $0.a_1a_2...$ where $a_i=0,1$.  Because the selected system is arbitrary given the translation symmetry of the space, we can consider our two sequences as indistinguishable if these real number representations match for any choice of the central element of each sequence.  But there are only countably many such choices, and uncountably many real numbers, so a given sequence is indeed 
distinguishable from some (indeed almost all) other sequences in this sense.

Now, {\em physically}, we can ask two key questions. First, is the difference between (finitely) {\em statistically indistinguishable} and {\em mathematically indistinguishable} important, given that any actual operation will only be able to gather a finite amount of statistical information? Second, given that these systems are by assumption identical and indistinguishable, and far outside of each others' horizon, is it meaningful to think of them as labeled (even if there {\em were} a preferred element in terms of their global distribution)?

If the answer to either question is negative, then it becomes unclear what purpose is served by distinguishing the elements in the post-measurement superposition, and one might ask whether in some future formulation of quantum cosmology they might be meaningfully identified, thereby rendering the issue wavefunction collapse fully irrelevant.
	
\subsection{The Chicken-and-Egg Problem of Quantum Spacetime}  
\label{ChickenEggSec}

We have argued that an infinite statistically uniform space can naturally emerge in modern cosmology, and place the quantum measurement problem in a very different light.  Yet quantum processes affect spacetime in any theory, and in inflation are responsible for the large-scale density fluctuations.  Moreover, some versions of eternal inflation themselves {\em rely} on quantum processes: stochastic eternal inflation is eternal solely due to quantum fluctuations of the inflaton, and in open eternal inflation, inflation {\em ends} due to quantum tunneling.  If quantum probabilities (and rates, etc.) are to be understood by making use of a cosmological backdrop, how do we make sense of the quantum processes involved in creating that backdrop?\footnote{Similar considerations might apply to other fields -- determined by quantum events but stretched into homogeneity by inflation -- that go into making up the ``background" in which the quantum experiment is posed, and to which the CP applies.}  There are at least four alternative ways in which we might view this chicken-and-egg problem. 

First, we might from an Everettian perspective consider such processes as simply parts of the unitary evolution (or non-evolution, if considering the Wheeler-DeWitt equation) of a wave-function(al).  Processes such as measurement of quantum systems by apparatuses and observers would not be meaningful until later times at which a classical approximation of spacetime had already emerged to describe an infinite space, in which such quantum measurement outcomes can be accorded probabilities via the arguments of this paper. How this view connects with the existing body of work on the emergence of classicality is an interesting avenue for further research.

Second, we might retain the Born rule as a axiomatic assumption, and employ it to describe such processes; then, for quantum measurements that exist as part of a spatial collection, the Born rule would be superfluous, as the probabilities for measurement outcomes are better considered as relative frequencies as per the arguments above.  

Third, we might consider a classical spacetime description as logically prior to the quantum one, and in particular postulate certain symmetries that would govern the spacetime in the classical limit. In this perspective we could simply assume an FLRW space obeying the CP \cite{Layzer}. Or, we might include inflation and eternal inflation, but postulate an appropriately generalized inflationary version of the CP (or {\em perfect} cosmological principle) governing the semi-classical universe (see ~\cite{Aguirre02,Aguirre03} for ideas along these lines).  Within this context, quantum events such as bubble nucleations could also be considered as part of a spatial collection of identical regions, and the whole set of arguments given herein could be applied to accord them probabilities.  This would, in the language of ~\cite{multiverse}, unify the ``Level II" (inflationary) multiverse with the ``Level III" (quantum) multiverse.

Fourth, we might imagine that a full theory of quantum gravity in some way fundamentally changes the quantum measurement problem, and that the considerations herein, based on standard quantum theory, apply only processes in a well-defined background spacetime.

This subtle issue is similar to the -- possibly related -- matter of Mach's principle in General Relativity (GR): applications of GR almost invariably implicitly assume a background frame that is more-or-less unaccelerated with respect to the material contents of the application; yet this coincidence between local inertial frames and the large-scale bulk distribution of matter is almost certainly of cosmological origin. Should Mach's principle simply be assumed as convenient, or explained as emerging in a particular limit from dynamics that do not assume it, or does it require specification of cosmological boundary conditions, or must new physics beyond GR be introduced? 
        
\section{Conclusions}
\label{ConcSec}

      Modern inflationary cosmology suggests that we exist inside an infinite statistically uniform space. 
If so, then any given finite system is replicated an infinite number of times throughout this space.  This raises serious conceptual issues for a prototypical measurement of a quantum system by an observer, because the measurer cannot know which of the identical copies she is, and must therefore ascribe a probability to each one \cite{Page09,Page09b,Page10,HartleSrednicki}. Moreover, as shown by Page \cite{Page09}, this cannot be seamlessly done using the standard projection operator and Born rule formalism of quantum mechanics; rather, it implies that quantum probabilities must be augmented by probabilities based on relative frequencies, arising from a measure placed on the set of observers. We have addressed this issue head-on by suggesting that perhaps it is not observer-counting that should be avoided, but quantum probabilities that should {\em emerge} from the relative frequencies across the infinite set of observers that exist in our three-dimensional space.
      
      To make this link between quantum measurement and cosmology, we have built on the classic work concerning frequencies of 
outcomes in repeated quantum measurements \cite{Finkelstein,Hartle68,Farhi89,Gutmann95}.
Our goal has not been to add further mathematical rigor (see \cite{CavesSchack94,Landsman,VanWesep,Wada07} for the 
current 
state-of-the-art\footnote{The core mathematical question is how to deal with the measure-zero set of confused freak observers; but as emphasized in \cite{everett3}, this issue is not unique to quantum mechanics, but occurs also in classical statistical mechanics and virtually other theory involving infinite ensembles.}),
but instead to develop these ideas in the new context of the physically real, spatial collection provided by cosmology.  
The argument shows that the product state of infinitely many existing copies of a quantum system can be rewritten as an infinite superposition of terms.  Because each term has zero norm, these must be grouped in terms of what they predict. Projecting these states with a ``confusion operator" shows that a grouping with total Hilbert space norm unity consists of terms all of which are functionally indistinguishable, and contain relative frequencies of measurement outcomes in precise accordance with the standard Born rule. The remaining states, which would yield different relative frequencies, have total Hilbert space norm zero.  
      
      Predictions of measurement outcomes probabilities in this situation are, then, provided entirely by relative frequencies; if conventional quantum probabilities enter at all, it is only to justify the neglect of the zero-norm portion of the global wavefunction. Because all terms in the unity-norm portion are indistinguishable, quantum interpretation must be done in a {\em cosmological} light.  Any ``collapse" of the wavefunction is essentially irrelevant, since collapse to any of the wavevectors corresponds to exactly the same outcome. In Everettian terms, the ``many worlds" are all the same; moreover, frequentist statistics emerge even for a single quantum measurement\footnote{For example, the author ordering for this paper was determined by a single quantum measurement, and the order you yourself read is shared by exactly half of all otherwise-indistinguishable worlds spread throughout space.}
(rather than a hypothetical infinite sequence of them). In this ``cosmological interpretation" of quantum mechanics, then, quantum uncertainty ultimately derives from uncertainty as to which of many identical systems the observer actually is.  
     
In conclusion, the quantum measurement issue is fraught with subtlety and beset by controversy; similarly, infinite and perhaps diverse cosmological spaces raise a host of perplexing questions and potential problems. We suggest here that perhaps combining these problems results not in a multiplication of the problems, but rather an elegant simplification in which quantum probabilities are unified with spatial observer frequencies, and the same infinite, homogeneous space that provides a real, physical collection also provides a natural measure with which to count. Further comprehensive development of this quantum-cosmological unification might raise further questions, but we hope that it may unravel further theoretical knots at the foundations of physics as well.

\bigskip
{\bf Acknowledgements:}
The authors are indebted and grateful to David Layzer for providing inspiration and copious insight, assistance, and feedback during the preparation of this paper. Andy Albrecht and Don Page provided very helpful comments. 
This work was supported by NASA grants NAG5-11099 and NNG 05G40G,
NSF grants AST-0607597, AST-0708534, AST-0908848, PHY-0855425, and PHY-0757912, a
``Foundational Questions in Physics and Cosmology" grant from the John Templeton Foundation,
and fellowships from the David and Lucile
Packard Foundation and the Research Corporation.   

\appendix

\section{Infinite statistically uniform spaces from eternal inflation}
\label{InflationDetailsSec}

In this appendix, we describe in greater detail how eternal inflation produces infinite spaces. Open inflation is perhaps the most well-studied case.  The Coleman-DeLuccia instanton~\cite{cdl} describes the spacetime and field configuration resulting from the nucleation of a single bubble where the inflaton field has lower energy.  In single-field models, each constant-field surface is a space of constant negative curvature, so that the bubble interior can be precisely described as an open Friedmann universe~\cite{cdl}.  Bubble collisions (which are inevitable) complicate this picture; see~\cite{bubblereview} for a detailed review. In no case, however, do collisions prevent the existence inside the bubble of an infinite, connected, spatial region\footnote{This region could be delimited by, e.g. defining some criterion by which to identify regions affected by the collision, and excluding them.  After this removal, the remaining region would have uniform properties (as would the statistics of the excisions; see below.)} 

	In topological eternal inflation~\cite{Vilenkin94,Linde94a}, a region of inflating spacetime is maintained by a topological obstruction such as a domain wall or monopole, where the field is at a local maximum of its potential.
The causal structure of these models is fairly well-understood if a bit subtle (see, e.g.,~\cite{Vilenkin94,Sakaietal00}), and also contains infinite spacelike surfaces after inflation has reheating surfaces.
	
	The precise structure of post-inflation equal-field surfaces in stochastic eternal inflation is less well-defined, as the spacetimes cannot be reliably calculated with analytic or simple numerical calculations.  However, they are expected to be infinite (e.g.,~\cite{winitz01,GarrigaVilenkin01}), with volume dominated by regions in which inflation has emerged naturally from slow roll~\cite{gratton10}, and thus space is relatively flat and uniform.  
	
	\begin{figure}[ht]
\centerline{\includegraphics[scale=0.3]{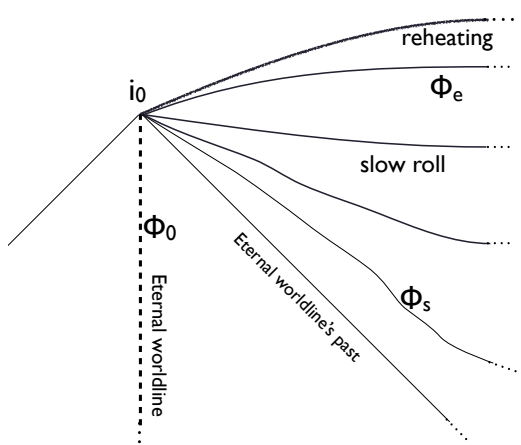}}
\caption{\label{EternalFig}
Generic conformal structure of eternal inflation and post-inflationary reheating surface in open, topological, or stochastic eternal inflation.
  A region in which inflation is eternal (i.e. contains timelike worldlines that can extend to infinite proper time while remaining in the inflation region) is bounded by an infinite surface of field value $\phi_s$ at which eternality fails.  Between this value and $\phi_e$ lie some number of e-folds of slow-roll inflation.  This slow-roll inflation ends on a spacelike surface that represents a natural equal-time surface for the subsequent evolution including reheating, etc.  This surface is spatially infinite, with spatial infinity denoted by $i_0$. (The future infinities following the reheating surface are not depicted). 
}
\end{figure}
	
	That infinite spacelike reheating surfaces are generic in eternal inflation is not surprising, as per the following heuristic argument.  Roughly speaking, eternal inflation occurs when some obstruction prevents the inflaton field $\phi$ from evolving away from some value $\phi_0$ during a typical Hubble time, so that on average, the physical volume containing that field value increases.  This implies that there are some (extremely rare) worldlines threading a horizon volume in which $\phi \approx \phi_0$ forever.  For such a model to be observationally viable, however, there must exist a route through field-space that crosses a value $\phi_s$ at which slow-roll inflation begins, then a value $\phi_e$ at which inflation ends and matter or radiation-domination begins.
	
	This field evolution plays out in spacetime as well, connecting the eternally inflating spacetime region to the post-inflationary region, as sketched in Fig.~\ref{EternalFig} for a single inflaton field.  Whether eternal inflation is open, topological, or stochastic, this diagram must look essentially the same.  The surface of constant field $\phi_s$, at which eternality fails, is by definition one for which very few worldlines cross back to field values near $\phi_0$; this surface must therefore be spacelike nearly everywhere.  Morever, slow-roll inflation to the future of this surface exponentially suppresses field gradients, so that surfaces of constant field quickly become uniformly spacelike\footnote{For a more precise specification of this point, see~\cite{VilenkinVanchurin,Sakaietal00}.} as the field approaches $\phi_e.$   
		
	Now, if we consider a spatial region with $\phi = \phi_e$ (soon to evolve to the reheating surface), and follow the surface of constant field $\phi=\phi_s$ in a direction toward the eternal region, we see three things.  First, the $\phi=\phi_e$ surface must be infinite, since the $\phi=\phi_s$ surface, for example, is infinite, and further inflates into the constant field surfaces with $\phi_s \le \phi \le\phi_e$.  Second, by the above argument the $\phi=\phi_e$ surface is {\em spacelike}, so there is no obstacle to continuing the foliation in our original region as far as we like toward the eternal region. Third, since each point on our surface has roughly the same classical field history since $\phi_s$ (variations due to the initial field velocity being bounded by the slow-roll condition, and curvature being stretched away), the $\phi=\phi_e$ surface should be uniform up to variations induced by quantum fluctuations during the field's evolution.
	
Multi-field models are more complex, in that we might imagine many routes for the vector of field valued $\vec \phi$ to take from a given (set of) `eternal' field values $\vec \phi_0$ to field values $\vec \phi_e$ at which inflation ends.  Yet it seems likely that in this case the prime difference would be for the $\phi = \phi_s$ surface to be replaced by an infinite, spacelike surface $\Sigma_s$ on on which eternality failes, and on which $\vec \phi$ is inhomogeneous.\footnote{This is precisely what happens in ``Quasi-open" inflation~\cite{Bellido+98}.  There, the bubble interior cannot be foliated into constant-field equal-time surfaces.  However, there is still an infinite space with infinitely many finite-sized regions in which the field is constant; a subset of these with the same field value could be taken as a statistically uniform (though potentially disconnected) space.} Yet for a given value $\vec\phi_s$ of this vector on this surface, subsequent evolution would be classically similar, so the subset $\Sigma_{\vec\phi_s}$ of $\Sigma_s$ containing this same vector of field values would evolve to a set $\sigma_{\vec\phi_s}$ of potentially disconnected regions\footnote{If the regions are disconnected, it becomes less obvious how to place a natural measure on the spatial volumes in them, but in some cases it may not make a significant difference. For example, if the spacetime can be foliated into surfaces of constant curvature, these will approximately coincide with the surfaces of constant field-value over an infinite domain sharing the same inflationary history, which could be taken as a statistically uniform space; but these slices would also provide a way to compute spatial volumes without ambiguity.} with uniform properties.  Because $\Sigma_s$ is infinite, but the fields on it can only take a finite range of values, determined randomly, it would seem that each $\sigma_{\vec\phi_s}$ must be infinite in volume as in the single field case.

The sense in which the post-inflationary spaces produced by inflation are statistically uniform, in the sense described in \Sec{ErgodicSec}, is rather subtle, as illustrated by the 
case of ``open" eternal inflation.  In an open FLRW space created by a single bubble nucleation in a background false-vacuum space, as described by~\cite{cdl}, the properties of the space are precisely uniform; this is guaranteed by the de Sitter symmetries obeyed by the progenitor space, which correspond to translations on a spatial slice within the bubble (see~\cite{bubblereview} and references therein for more detail.)  However, a single isolated bubble is not realistic, and we must consider potential bubble collisions. For a given worldline inside a bubble, collisions come in two classes: ``early" bubbles that enter the past of the worldline within its first Hubble time, and ``late bubbles" that enter later.  Early bubbles affect nearly all worldlines, and do {\em not} obey the CP: there is a distinguished position at which an observer sees the lowest rate of collisions.  This defines a frame that breaks the de Sitter symmetry of the background space into the group of rotations and spacetime translations. However, there exists an infinite set of wordlines that see {\em no} early collisions, and are surrounded by an exponentially large region that is also unaffected.  ``Late" bubbles do obey the CP in terms of their probabilities, and can affect a fraction of volume that can be either large or exponentially small.  In both early and late bubbles, a given type of collision may or may not significantly disrupt the dynamics of inflation within the region it affects, depending upon the details of the collision.  

The bottom line is that there exists an infinite spacetime with statistically uniform properties up to some large scale, corresponding to the typical distance between areas affected by bubbles.  On these larger scales, the properties of the bubble-collision regions themselves select a preferred position if ``early" bubbles are included, but are homogeneous if only ``late" bubbles are included.  In any case, an infinite set of regions larger than our observable universe, with randomly chosen initial conditions draw from the same probability distribution, exists.  

Is this by itself enough for the arguments of this paper to go through?  In principle, the {\em measure} by which the observers in this collection are counted can affect relative frequencies of observer types.  In simple cases, such as a pure FLRW universe as might arise in single bubble nucleation, the spatial rotation and translation symmetries pick out a preferred measure given by spatial volume.  For more complicated spaces in which these symmetries are less clear, the measure can be quite important for certain questions. However, this is only really true if the measure depends in some way on the relevant information discriminating these types of observers.  For questions of cosmological observations, this can easily be the case.   In our case of a local laboratory measurement, it is hard to see how the measure can depend on the outcome of a Stern-Gerlach experiment, thus hard to see how the measure could enter the relative frequencies of observed outcomes in the collection.

\section{Generalization to observations with more than two outcomes}
\label{GeneralizationSec}

In this section, we generalize the result that 
$\|\>\confusionoperator\>\psiket\|^2\to 0$ as $N\to\infty$
from 2-state systems to systems with an arbitrary number of states $m\ge 2$.
Instead of the two basis vectors $\down$ and $\up$, we now suppose $m$ orthogonal basis vectors labeled as 
$|\>1\rangle, 
|\>2\rangle,
\dots,
|\>m\rangle$.
Now consider a finite uniform space that is large enough to contain $N$ identical copies of our system
(which we will keep referring to as a ``particle'' for simplicity, even though our proof is valid for an arbitrary $m$-state system), each
prepared in the state
\beq{InitialStateEq}
|\psi\rangle = \alpha_1|\>1\rangle + \alpha_2|\>2\rangle + \dots + \alpha_m|\>m\rangle. 
\eeq
Analogously with \eq{PsiEq}, 
the state of this combined $N$-particle system is simply a tensor product with $N$ terms, each of which involves a 
sum of
$N$ basis vectors. 
The frequency operator $\freqoperator$ from above is naturally generalized to an $m$-dimensional vector operator
$\freqvecoperator$, whose $i^{\rm th}$ component $\freqvecoperator_i$
measures the frequency of the outcome $|\>i\rangle$. 
In other words, \eq{FrequencyOperatorxample} is diagonal in the 
$m^N$-dimensional basis spanned by products of $N$ of our 1-particle states, 
and the eigenvalues of $\freqvecoperator_i$ are the frequencies of $|\>i\rangle$ in these basis vectors. 
For example, for $N=5$ and $m=5$, 
\beq{vecfreqoperatorExampleEq}
\freqvecoperator |\>3\rangle |\>1\rangle |\>4\rangle |\>1\rangle |\>5\rangle  = 
\left(
\begin{tabular}{c}
$2/5$\\
$0$\\
$1/5$\\
$1/5$\\
$1/5$
\end{tabular}
\right)
|\>3\rangle |\>1\rangle |\>4\rangle |\>1\rangle |\>5\rangle.
\eeq
Examining the norm of the state multiplied by $(\freqvecoperator-\p)$, where the $m$-dimensional vector $\p$ is defined by 
$p_i\equiv|\alpha_i|^2$, 
\eq{FproofEq} now generalizes to
\beqa{FproofEq2}
& &\psibra(\freqvecoperator_i-p_i)(\freqvecoperator_j-p_j)^t\psiket=\nonumber\\
&=&\sum_{n_1\dots n_m}\left({N\atop n_1...n_m}\right) p_1^{n_1}\dots p_m^{n_m}\left({n_i\over N}-p_i\right)\left({n_j\over N}-p_j\right)=\nonumber\\
&=&\C_{ij}={\C_{ij}^*\over N^2},
\eeqa
where the sum on the second row is over all $m$-tuplets of natural numbers $(n_1,...,n_m)$ such that $\sum_{i=1}^m n_i=N$, and 
 the parenthesis following the summation symbol denotes the multinomial coefficient $N!/n_1!...n_m!$. The $m\times m$ matrix
$\C^*$ is given by
\beq{CdefEq}
\C^*_{ij}=\left\{
\begin{tabular}{cccccccc}
$Np_i(1-p_i)$ 	&if $i=j$,\\
$-Np_i p_j$	&if $i\ne j$.
\end{tabular}
\right.
\eeq
The terms $\left({N\atop n_1...n_m}\right) p_1^{n_1}\dots p_m^{n_m}$
in \eq{FproofEq2} are simply the familiar coefficients of a multinomial distribution that 
has mean vector $N\p$. 
We recognize the last two lines of \eq{FproofEq2} as the definition of the covariance matrix $\C^*$ of the multinomial distribution, and the proof of \eq{CdefEq} is simply the well-known multinomial covariance derivation.

Since $\C\equiv\C^*/N^2\propto 1/N\to 0$ as $N\to\infty$, the multinomial distribution governing the spread in values of the frequency operator vector in Eq.~\ref{FproofEq2} approaches an $m$-dimensional $\delta$-function centered on the vector $\p$.
In particular, the special case $i=j$ implies that 
\beq{FreqVecLimitEq}
\|(\freqvecoperator_i-p_i)\psiket\|^2  = p_i(1-p_i)/N \to 0,
\eeq
so that $\psiket$ becomes an eigenvector of all of the components of $\freqvecoperator$ (again in the sense that the correction term has zero norm), with eigenvalues corresponding to the coefficients of the vector $\p$, as shown in the seminal paper \cite{Hartle68}.

Let us now consider the confusion operator $\>\confusionoperator\>$. Since we are now considering not merely one frequency but $m$ different frequencies $p_i$, we generalize our definition of $\>\confusionoperator\>$ to be the projection operator onto 
those basis vectors where the spin-up fraction
differs by more than a small predetermined value $\epsilon$ from the 
Born rule prediction $p_i=|\beta_i^2|$ for {\it any} $i=1,...,m$.
It is therefore related to the frequency operator vector $\freqvecoperator$ by
\beq{OperatorRelationEqGen}
\confusionoperator = \Ihat - \prod_{i=1}^m\theta\left(\epsilon-|\freqoperator_i-p_i|\right),
\eeq
where $\theta$ again denotes the Heaviside step function and $\Ihat$ is the identity operator.

To gain intuition about the confusion operator $\confusionoperator$, it is helpful to consider the  $m$-dimensional generalization of \fig{BinomialFig}, the $m$-dimensional space 
of vectors $\f$ where the $i^{\rm th}$ component $f_i=n_i/N$ is the frequency of the outcome 
$|\>i\rangle$. In the basis where $\confusionoperator$ is diagonal
(with basis vectors like the example $|\>3\rangle |\>1\rangle |\>4\rangle |\>1\rangle |\>5\rangle$), 
each basis vector maps to a unique point $\f$ in this space which is the vector of eigenvalues exemplified in \eq{vecfreqoperatorExampleEq}, \ie, 
a point with coordinates corresponding to the various frequencies $f_i=n_i/N$.
In this space, the eigenvalues of $\confusionoperator$ equal $0$ for eigenvectors falling inside an $m$-dimensional hypercube of side length $2\epsilon$ centered at the point $\p$, and otherwise equal $1$.
Analogously to \eq{ConfusionProofEq}, 
we obtain
\beqa{ConfusionProofEq2}
&&\|\>\confusionoperator\>\psiket\|^2=\psibra\>\confusionoperator\>\psiket=\nonumber\\
&=&\sum_{n_1...n_m}\left({N\atop n_1...n_m}\right)p_1^{n_1}...p_m^{n_m}
\left[1 - \prod_{i=1}^m\theta\left(\epsilon-\left|{n_i\over N}-p_i\right|\right)\right]\nonumber\\
&=&\sum_{\hbox{\small\rm Outside cube}}\left({N\atop n_1...n_m}\right)p_1^{n_1}...p_m^{n_m}.
\eeqa
The sum on the second row is again over all $m$-tuplets sets of natural numbers $(n_1,...,n_m)$ such that $\sum_{i=1}^m n_i=N$,  
and the sum on the last row is restricted to the subset lying outside of the above-mentioned hypercube where all frequencies are within $\epsilon$ of 
the Born rule prediction.

Since the binomial distribution for the frequency vector $\f$ is known to approach a $\delta$-function $\delta(\f-\p)$ as $N\to\infty$, 
it is obvious from \eq{ConfusionProofEq2} that
$\|\>\confusionoperator\>\psiket\|^2\to 0$ in this limit.
Just like for the 1-particle case, the convergence will be quite rapid (faster than polynomial) once the width of the multinomial distribution starts becoming significantly smaller than 
that of the surrounding hypercube. This is because by the central limit theorem, the multinominal distribution becomes well-approximated by a Gaussian in the $N\to\infty$ limit, 
so we can approximate the sum in \eq{ConfusionProofEq2} by an integral over a multivariate Gaussian with mean $\p$ and covariance matrix $\C=\C^*/N^2$:
\begin{widetext}
\beqa{ConfusionProofEq3}
\|\>\confusionoperator\>\psiket\|^2 &=&\sum_{\hbox{\small\rm Outside cube}}\left({N\atop n_1...n_m}\right)p_1^{n_1}...p_m^{n_m}
 \approx \int_{\hbox{\small\rm Outside cube}}{1\over (2\pi)^{N/2}|\C|^{1/2}} e^{-{1\over 2}(\f-\p)^t\C^{-1}(\f-\p)}d^N f\nonumber\\
&\le&\int_{\hbox{\small\rm Outside cube}}{1\over (2\pi)^{N/2}|{\I\over 2N}|^{1/2}} e^{-{1\over 2}(\f-\p)^t({\I\over 2N})^{-1}(\f-\p)}d^N f = \int_{\hbox{\small\rm Outside cube}}{1\over (\pi/N)^{N/2}} e^{-N|\f-\p|^2}d^N f\nonumber\\
&=&1-\int_{\hbox{\small\rm Inside cube}}{1\over (\pi/N)^{N/2}} e^{-N|\f-\p|^2}d^N f = 1-\int_{\hbox{\small\rm Inside cube}}{1\over (\pi/N)^{N/2}} \prod_{i=1}^N e^{-N(f_i-p_i)^2} d^N f\nonumber\\
&=&1-\erf\left[\epsilon N^{1/2}\right]^N = 1-\left(1-\erfc\left[\epsilon N^{1/2}\right]\right)^N\nonumber\\
&\to&  N\erfc\left[\epsilon N^{1/2}\right]\to 0\quad\hbox{as }N\to\infty.
\eeqa
\end{widetext}
On the second line, we used the readily proven fact that no eigenvalue of $\C$ can exceed $1/2N$, 
which means that if we replace the covariance matrix $\C$ by $\I/2N$ (where $\I$ is the identity matrix), 
then the multivariate Gaussian remains at least as wide in all directions, meaning that the fraction of its
integral residing outside of the hypercube is at least as large.

\section{Measurement by the collection of macroscopically indistinguishable apparatuses}
\label{RhoAppendix}

In \Sec{DecohereSec} we treated the case in which every measuring apparatus was in an identical quantum state.
Let us now generalize this to the arguably more relevant general case where the initial state of the apparatus is described by a density matrix.
Specifically, we can write the density matrix of the apparatus in the 
``ready" state as
\beq{ApparatusRhoEq}
\rho_a = \sum_i p_i \iready\ireadybra,
\eeq
where $i$ indexes the 
``ready'' microstates $\iready$ of the apparatus
that are macroscopically indistinguishable, and the coefficients $p_i$ satisfy 
$p_i\ge 0$ and $\sum_i p_i =1$.\footnote{The density matrix $\rho_a$ can without loss of generality be written in the diagonal form of \eq{ApparatusRhoEq}, because if it were not diagonal, then we could make it diagonal by defining new ``ready'' basis states that are the eigenvectors of $\rho_a$; they form an orthogonal basis because $\rho_a$ is Hermitean, and they behave like classical apparatus states because they are superpositions of classical apparatus states (the old basis states) that are macroscopically indistinguishable.
}
(The coefficients $p_i$ are commonly interpreted as the probability of finding the apparatus in states $\iup$, but no such interpretation is needed for our argument below.)
The initial density matrix for the combined system and apparatus is thus
\beq{TotalInitialRhoEq}
\rho=\sum_i p_i(\alpha\up + \beta\down)\iready \ireadybra(\alpha^*\upbra + \beta^*\downbra).
\eeq

With this notation, the unitary evolution during the measurement process 
given by \eq{vnpremeasure} takes the form 
\beqa{vnpremeasure2}
(\alpha\up + \beta\down)\iready \longrightarrow \alpha\up\iup + \beta\down\idown,
\eeqa
where $\iup$ and $\idown$ are the corresponding microstates of the apparatus in which it records an `up' or `down' measurement. 
We thus have two resulting classes of apparatus states $\{\iup\}$ and $\{\idown\}$, where the states in each class are macroscopically indistinguishable, but where the two classes are macroscopically distinguishable (having a macroscopic pointer in different locations, say). 

Let us now consider what happens to the product state of all the apparatuses and systems.
The initial product state of \eq{PsiEqApp} generalizes to a tensor product of $N$ density matrices
that are all given by \eq{TotalInitialRhoEq}, so each of the $N$ factors contains 
a sum over the local apparatus microstates $i$. If we expand this product of sums, then this total density matrix
takes the form of a weighted average 
\beq{bigrho}
\rho = \sum_{i_1,...,i_N} p_{i_1}...p_{i_N}\rho_{i_1,...i_N},
\eeq
over all possible combinations of apparatus microstates. Here, $i_1...i_N$ each run over the set of detector states, $p_{i}$ are as before, and $\rho_{i_1,...,i_N}$ are
 pure-state density matrices of the form 
$\psiket\psibra$, where $\psiket$ is a product of the form
\beq{InitialComponentEq}
\psiket = \up|17_r\rangle \tensormult\up|4711_r\rangle \tensormult\down|5_r\rangle\tensormult\up|17_r\rangle \tensormult\down|666_r\rangle...
\eeq
(for example, $|666_r\rangle$ denotes the apparatus ``ready" microstate $\iready$ with $i=666$).
Now, after the interaction between the system and apparatus described by \eq{vnpremeasure2}, 
each of these pure-state density matrices evolves into a new $\rho'_{i_1,...i_N}=|\psi'\rangle\langle\psi'|$, where $|\psi'\rangle$ is a product of the form
\beq{FinalComponentEq}
\psiket = \up|17_\uparrow\rangle \tensormult\up|4711_\uparrow\rangle \tensormult\down|5_\downarrow\rangle\tensormult\up|17_\uparrow\rangle \tensormult\down|666_\downarrow\rangle...
\eeq
We see that the only difference between these pure states $|\psi'\rangle$
and the ones we considered in \eq{PsiEqInfApp} above is that the apparatus microstates now vary in a typically random-looking fashion (yet always in such a way that the an $\up$ particle state goes with one of the apparatus microstates in the 
``up'' class). 
Applying the exact same reasoning as  \Sec{FiniteSec} and \Sec{InfiniteSec} to one of these pure states $|\psi'\rangle$
when $N\to\infty$, our product state therefore becomes an infinite superposition of terms like \eq{FinalComponentEq},
all of which (except for a set of total Hilbert space norm zero) 
have the relative frequencies $|\alpha|^2$ and $|\beta|^2$ for terms with apparatus states in the ``up'' and ``down'' classes,
respectively.  
Since this result holds for each $\rho'_{i_1,...,i_N}$, it clearly applies to $\rho$: 
it describes an collection of apparatuses, a fraction $|\alpha|^2$ of which measure ``up'' and a fraction 
$|\beta|^2$ of which measure ``down'' (up to a part of zero Hilbert space norm as usual).

Note that this density matrix and the one described in \Sec{DecohereSec}
have a different structure; but we can combine them if we also consider the interaction of our set of measuring devices with their local environments.  In this case, if there are $N$ particles, then after tracing out the environment, we have an overall density matrix that looks just like \Eq{bigrho}, but in which each of the $\rho_{i_1,...,i_N}$ is now a diagonal density matrix 
of just the type described in \Sec{DecohereSec}, except that each apparatus is labeled by its state $i_k$, and all such combinations are combined as a weighted sum to get the total density matrix.

%%%%%%%%%%%%%%%%%%%%%%%%%%%%%%%%%%%%%%%%%%%%%%%%%%%%%%%%%%%%
%%%%%%%%%%%%%%%%%%%%%% REFERENCES: %%%%%%%%%%%%%%%%%%%%%%%%%

\bibliographystyle{ieeetr}

\begin{thebibliography}{99}
%\section{References} 
%\begin{references}   %
% EVERETT QM:

\bibitem{Jammer} \rfbook\nn Jammer M;1974;The Philosophy of Quantum Mechanics: The Interpretations of Quantum Mechanics in Historical Perspective;New York;Wiley \& Sons.

\bibitem{Born26}
% Birth of the Born Rule
\rf\nn Born M;1926;Z.Phys.;37;863
% Zur Quantenmechanik der Stoßvorgänge, Max Born, Zeitschrift für Physik, 37, #12 (Dec. 1926), pp. 863-867 (German); English translation in Quantum theory and measurement, section I.2, J. A. Wheeler and W. H. Zurek, eds., Princeton, NJ: Princeton University Press, 1983.

\bibitem{Ballantine}
\rfbook\nn Ballantine L;1998;Quantum Mechanics: A Modern Development;Singapore;World Scientific Publishing Company

% COPENHAGEN INTERPRETATION:
\bibitem{Bohr}
\rfproc\nn Bohr N;1985;Collected Works;\nn Kalchar J;North-Holland;Amsterdam

\bibitem{Heisenberg}
\rfproc\nn Heisenberg W;1989;Collected Works;\nn Rechenberg W;Springer;Berlin
% Title:	Collected works
% Author(s):	 Werner Heisenberg.
% Date:	1989
% Publisher:	Springer-Verlag
% Pages:	717
% Series:	Original Scientific Papers, Series A, Part II
% Note:	German language
% Subject:	 Physics , Heisenberg, Werner
% Call #:	QC23:H44:1989 Take a Virtual Look at the Shelves
% Availability:	IN     Request a Copy

\bibitem{Dirac} \rfbook\nnn Dirac P A;1967;The Principles of Quantum Mechanics, 4th ed., revised;Oxford;{Clarendon Press}

\bibitem{Wigner}
\rn\nnn Wigner E P 1976, Princeton University Lecture Notes on Interpretation of Quantum Mechanics (unpublished).

\bibitem{WheelerZurek}
\rfbook\nnn Wheeler, J A \dualand \nnn Zurek W H;1983;Quantum Theory and Measurement;Princeton University Press;Princeton

\bibitem{GottfriedYan}
\rfbook\nn Gottfried K \dualand Yan T;2003;Quantum Mechanics: Fundamentals;Springer;Berlin


% HYDRODYNAMIC INTERPRETATION:
\bibitem{Madelung27}
\rf\nn Madelung E;1927;Z.Phys.;40;322
% Title:	
% Quantentheorie in hydrodynamischer Form
% Authors:	
% Madelung, E.
% Publication:	
% Zeitschrift für Physik, Volume 40, Issue 3-4, pp. 322-326
% Publication Date:	
% 03/1927
% Origin:	
% SPRINGER
% DOI:	
% 10.1007/BF01400372
% Bibliographic Code:	
% 1927ZPhy...40..322M

% CONSCIOUSNESS INTERPRETATION:
  % STANFORD ENCYPLOPEDIA OF PHILOSOPHY:
  % By contrast to von Neumann's fairly cautious stance, London and Bauer (1939) went much further and proposed that it is indeed human consciousness which completes quantum measurement (see Jammer (1974, Sec. 11.3 or Shimony (1963) for a detailed account). In this way, they attributed a crucial role to consciousness in understanding quantum measurement - a truly radical position. In the 1960s, Wigner (1967) followed up on this proposal,[7] coining his now proverbial example of "Wigner's friend". In order to describe measurement as a real dynamical process generating irreversible facts, Wigner called for some nonlinear modification of (2) to replace von Neumann's projection (1).[8]

\bibitem{Hoeffding} \rf \nn Hoeffding W;1963;Journal of the American Statistical Association;58 (301);13Ð30

\bibitem{vonNeumann32}
\rfbook\nn {von Neumann} J.;1932;Die mathematischen Grundlagen der Quantenmechanik;Springer;Berlin.

\bibitem{LondonBauer39}
\rfproc\nn London F\dualand\nn Bauer E;1939;Quantum Theory and Measurement (1983);\nnn Wheeler J A\dualand\nnn Zurek W H;{Princeton Univ.~Press};Princeton
% London, F., and Bauer, E. (1939). La théorie de l'observation en mécanique quantique. Hermann, Paris. English  translation: The theory of observation in quantum mechanics. In Quantum Theory and Measurement, ed. by J.A. Wheeler and W.H. Zurek, Princeton University Press, Princeton, 1983, pp. 217-259.

\bibitem{Wigner67}
\rfbook\nnn Wigner E P;1967;Symmetries and Reflections;{Indiana University Press};Bloomington
% Remarks on the mind-body question. In Symmetries and Reflections, Indiana University Press, Bloomington, pp. 171-184.

\bibitem{Stapp93}
\rfbook\nnn Stapp H P;1993;In Mind, Matter, and Quantum Mechanics;Springer;Berlin
% Stapp, H.P. (1993). "A quantum theory of the mind-brain interface". In Mind, Matter, and Quantum Mechanics, Springer, Berlin, pp. 145-172.


% PILOT WAVE INTERPRETATION:
% de Broglie, L., 1928, in Solvay 1928 - BUT NOTHING PUBLISHED, IT SEEMS!
\bibitem{Bohm52}
\rf\nn Bohm D;1952;Phys.Rev.;85;166
% Bohm, David (1952). "A Suggested Interpretation of the Quantum Theory in Terms of "Hidden Variables" I". Physical Review 85: 166-179. doi:10.1103/PhysRev.85.166.

% QUANTUM LOGIC:
\bibitem{BirkhoffNeumann36}
\rf\nn Birkhoff G\dualand\nn {von Neumann} J;1936;Ann.Math.;37;823
% G. Birkhoff and J. von Neumann, The Logic of Quantum Mechanics, Annals of Mathematics, Vol. 37, pp. 823-843, 1936.

% EVERETT:
\bibitem{Everett57}
\rf\nn Everett H;1957;Rev. Mod. Phys;29;454

\bibitem{EverettBook}
\rfproc\nn Everett H;1957;The Many-Worlds Interpretation of Quantum 
Mechanics;\nnn DeWitt B S\dualand\nn Graham N;Princeton;{Princeton Univ. Press,
available at \url{http://www.pbs.org/wgbh/nova/manyworlds/pdf/dissertation.pdf}}

% STOCHASTIC MECHANICS:
\bibitem{Nelson66}
\rf\nn Nelson E;1966;Phys.Rev.;150;1079
%  Nelson,E. (1966) Derivation of the Schrödinger Equation from Newtonian Mechanics, Phys. Rev. 150, 1079-1085

% MANY MINDS:
\bibitem{Zeh70}
\rf\nnn Zeh H D;1970;Found.Phys.;1;69
% On the interpretation of measurement in quantum theory", 1970, Foundations of Physics, Volume 1, Issue 1, pp.69-76

\bibitem{Page}
\rf\nnn Page D N;1996;Int.J.Mod.Phys.;D5;583
% Sensible Quantum Mechanics: Are Probabilities only in the Mind?
% Don N. Page (University of Alberta, Edmonton, Canada)
% Comments:	LaTeX, 14 pages, shortened version of quant-ph/9506010
% Subjects:	General Relativity and Quantum Cosmology (gr-qc); Quantum Physics (quant-ph)
% Journal reference:	Int.J.Mod.Phys. D5 (1996) 583-596
% Report number:	Alberta-Thy-13-95
% Cite as:	arXiv:gr-qc/9507024v1

\bibitem{Griffiths84}
\rf\nnn Griffiths R B;1984;J.Stat.Phys.;36;219
% Title:	
% Consistent histories and the interpretation of quantum mechanics
% Authors:	
% Griffiths, Robert B.
% Affiliation:	
% AA(Department of Physics, Carnegie-Mellon University; , Institut des Hautes Etudes Scientifiques)
% Publication:	
% Journal of Statistical Physics, Volume 36, Issue 1-2, pp. 219-272
% Publication Date:	
% 07/1984

% OBJECTIVE COLLAPSE INTERPRETATION:
\bibitem{GRW86}
\rf\nn Ghirardi G C, \nn Rimini A\multiand\nn Weber T;1986;PRD;34;470

% TRANSACTIONAL INTERPRETATION:
\bibitem{Cramer86}
\rf\nn Cramer J;1986;Rev.Mod.Phys.;58;647
% The Transactional Interpretation of Quantum Mechanics by John Cramer. Reviews of Modern Physics 58, 647-688, July (1986)

% MODAL INTERPRETATION:
\bibitem{vanFraassen72}
\rfproc\nn {van Fraassen} B;1972;Paradigms and Paradoxes: The Philosophical Challenge of the Quantum Domain;
\nn Colodny R;{Univ.~Pittsburgh Press};Pittsburgh
% Bas van Fraassen, 1972, "A formal approach to the philosophy of science," in R. Colodny, ed., Paradigms and Paradoxes: The Philosophical Challenge of the Quantum Domain. Univ. of Pittsburgh Press: 303-66.

% EXISTENTIAL INTERPRETATION:
\bibitem{Zurek87}
\rfprep\nnn Zurek W H;1987;arXiv:0707.2832
% Relative States and the Environment: Einselection, Envariance, Quantum Darwinism, and the Existential Interpretation
%
% Decoherence and the existential interpretation of quantum theory, or "no information without representation," pp. 341-350 in From Statistical Physics to Statistical Inference and Back, P. Grassberger and J.-P. Nadal, eds. (Kluwer,  Dordrecht, 1994)

% RELATIONAL INTERPRETATION:
\bibitem{Rovelli96}
\rf\nn Rovelli C;1996;Int. J. of Theor. Phys.;35;1637
% Carlo Rovelli, 1996, "Relational Quantum Mechanics," Int. J. of Theor. Phys. 35: 1637. Also arXiv: quant-ph/9609002

% MONTEVIDEO INTERPRETATION:
\bibitem{GambiniPullin09}
\rfprep\nn Gambimi R\dualand\nn Pullin J;2009;arXiv:0903.1859
% Free will, undecidability, and the problem of time in quantum gravity
% Rodolfo Gambini, Jorge Pullin
% (Submitted on 10 Mar 2009)
% Subjects:	Quantum Physics (quant-ph); General Relativity and Quantum Cosmology (gr-qc); High Energy Physics - % Cite as:	arXiv:0903.1859v1 [quant-ph]


% DECOHERENCE:

\bibitem{bohm} \rn D. Bohm, Quantum Theory, 1950 (1989 Dover edition), Ch. 22

% Zeh (above)

\bibitem{Zurek09}
\rf\nnn Zurek W H;2009;Nature;5;181
% arXiv:0903.5082
% Quantum Darwinism
% Wojciech Hubert Zurek
% Journal-ref: Nature Physics, vol. 5, pp. 181-188 (2009)
% Subjects: Quantum Physics (quant-ph)

\bibitem{JZ85}
\rf\nn Joos E\dualand\nnn Zeh H D;1985;Z. Phys. B;59;223

\bibitem{SchlosshauerBook}
\rfbook\nn Schlosshauer M;2007;Decoherence and the Quantum-To-Classical Transition;Springer;Berlin


% LEVEL I MULTIVERSE:
\bibitem{GarrigaVilenkin01}
\rf\nn Garriga J\dualand\nn Vilenkin A;2001b;Phys. Rev. D;64;043511.
% SPACE IS INFINITE
% arXiv:gr-qc/0102010 [pdf, ps, other]
% Many worlds in one
% Jaume Garriga, Alexander Vilenkin
% Comments: 9 pages, 2 figures, comments and references added
% Journal-ref: Phys.Rev. D64 (2001) 043511



%%% PAGE BORN RULE DIES:

\bibitem{Page09}
\rf\nnn Page D N;2009;JCAP;0907;008
% arXiv:0903.4888 [pdf, ps, other]
% The Born Rule Dies
% Don N. Page
% Comments: LaTeX, 16 pages, typos in Eqs. (4.3) and (6.2) corrected
% Journal-ref: JCAP 0907:008,2009
% Subjects: High Energy Physics - Theory (hep-th); General Relativity and Quantum Cosmology (gr-qc); Quantum Physics % (quant-ph)


\bibitem{Page09b}
\rfprep\nnn Page D N;2009;arXiv:0907.4152
% arXiv:0907.4152 [pdf, ps, other]
% Born Again
% Don N. Page
% Comments: 4 pages, LaTeX
% Subjects: High Energy Physics - Theory (hep-th); Quantum Physics (quant-ph)

\bibitem{Page10}
\rfprep\nnn Page D N;2010;arXiv:1003.2419
% arXiv:1003.2419 [pdf, ps, other]
% Born's Rule Is Insufficient in a Large Universe
% Don N. Page
% Comments: LaTeX, 7 pages
% Subjects: High Energy Physics - Theory (hep-th); General Relativity and Quantum Cosmology (gr-qc); Quantum Physics 

%%% THE CLASSIC FREQUENCY OPERATOR PAPERS:
\bibitem{Finkelstein}
\rf\nn Finkelstein D;1963;Trans.N.Y.Acad.Sci;25;621
% APPARENTLY REPRINTED IN THE EVERETT BOOK!

\bibitem{Hartle68}
\rf\nn Hartle J;1968;Am.J.Phys.;36;704
% Quantum meachanics of individual systems

\bibitem{HartleSrednicki} 
\rf\nn Hartle J, and \nn Srednicki M;2010;PRD;81;123524

\bibitem{Farhi89}
\rf\nn Farhi E, \nn Goldstone J\multiand\nn Gutmann S;1989;Ann.Phys.;192;368

\bibitem{Gutmann95}
\rf\nn Gutmann S;1995;PRA;52;3560



\bibitem{LayzerBook} \rfbook \nn Layzer D;1990;Cosmogenesis;Oxford University Press;

\bibitem{Layzer1} \nn Layzer D 1971, ``Cosmogonic Processes in Astrophysics 
and General Relativity'', in {\it Brandeis Summer School Lectures 1968}, 
ed. M. Chretien, S. Deser, and J. Goldstein

\bibitem{Layzer}
\rfprep  \nn Layzer D;2010; ArXiv:1008.1229


\bibitem{vilenkinreplicas}
\rf \nn Knobe J, \nnn  Olum K D, \nn Vilenkin A;2006;{British Journal for the Philosophy of Science};57;47

\bibitem{guth}
\rf \nn Guth A;1981;PRD;23;347

\bibitem{LindeRev}
\rf\nn Linde A;2008;Lect.Notes Phys.;738;1

\bibitem{GuthRev}
\rf\nn Guth A;2007;J. Phys. A;40;6811

 \bibitem{AguirreRev}
 \rfprep \nn Aguirre A;2008;ArXiv:0705.0164

\bibitem{Lindehybridnoteternal} \rf \nn Linde A;2005;Physica Scripta;40;

\bibitem{wands} 
\rf \nnn Copeland E J, \nnn Liddle A R, \nnn Lyth D H,\nnn Stewart E D\multiand\nn Wands D;1994;PRD;49;6410

%\bibitem{wmap7}\rf \nn Komatsu E;et al.;2010;ArXiv;arXiv:1001.4538


\bibitem{wmap7}
\rfprep\nn Komatsu E {\etal};2010;arXiv:1001.4538
% Seven-Year Wilkinson Microwave Anisotropy Probe (WMAP) Observations: Cosmological Interpretation
% E. Komatsu, K. M. Smith, J. Dunkley, C. L. Bennett, B. Gold, G. Hinshaw, N. Jarosik, D. Larson, M. R. Nolta, L. Page, D. N. Spergel, M. Halpern, R. S. Hill, A. Kogut, M. Limon, S. S. % Meyer, N. Odegard, G. S. Tucker, J. L. Weiland, E. Wollack, E. L. Wright
% Comments: 51 pages, 19 figures. Submitted to Astrophysical Journal Supplement Series. (v2) References added. The SZ section expanded with more analysis. 
% The discrepancy between the KS and X-ray derived profiles has been resolved

\bibitem{tegmarkkachru} \rf\nnn Hertzberg M P, \nn Tegmark M, \nn Kachru S, \nn Shelton J, Ozcan O;2007;PRD;76;103521

\bibitem{bdg} \rf\nn Banks T, \nn Dine M, \nn Gorbatov, E;2004;JHEP;0408;058

\bibitem{arkaniswamp} \rf\nn Arkani-Hamed N, \nn Dubovsky S, \nn Nicolis A, \nn Trincherini E, \nn Villadoro G; 2007;JHEP;0705;055

\bibitem{banksfinite} \rfprep \nn Bankt T, \nn Johnson M;2005;arXiv:hep-th/0512141

\bibitem{winitskireview}
\rf\nn Winitzki, S;2008;Lect.Notes~Phys.;738;157

\bibitem{sdsspars}
% Cosmological parameters from SDSS and WMAP
\rf\nn Tegmark M {\etal};2004;PRD;69;103501	
% astro-ph/0310723

\bibitem{VN32} \rn J. von Neumann, 1932, {\em Mathematische Grundlagen der Quantenmechanik} (Springer, Berlin)

\bibitem{collapse}
\rf\nn Tegmark M;1993;Found.~Phys.~Lett.;6;571

\bibitem{brain}
\rf\nn Tegmark M;2000;PRE;61;4194
% The importance of quantum decoherence in brain processes;M. Tegmark 2000;Phys. Rev. E;61;4194-4206 
% quant-ph/9907009

\bibitem{multiverse}
\rf\nn Tegmark M;2003;Sci.~Am.;270 (5/2003);40
% Parallel Universes
\bibitem{multiverse4wheeler}
\rfprep\nn Tegmark M;2003;astro-ph/0302131


\bibitem{Aguirre03}
\rf\nn Aguirre A\dualand\nn Gratton S;2003;Phys.Rev. D;67;083515

\bibitem{Aguirre02}
\rf\nn Aguirre A\dualand\nn Gratton S;2002;Phys.Rev. D;65;083507


\bibitem{CavesSchack94}
\rf\nnn Caves C M\dualand\nn Schack R;2005;Ann.Phys.;315;123
% C.M. Caves and R. Schack, Ann. Phys. 315, 123--146 (2005)
% Properties of the frequency operator do not imply the quantum probability postulate
% Carlton M. Caves, Ruediger Schack

\bibitem{Landsman}
\rfprep\nnn Landsman N P;2008;arXiv:0804.4849v1



\bibitem{VanWesep}
\rf\nnn Van Wesep R A; 2006;Annals of Physics;321;2438

\bibitem{Wada07}
\rf\nn Wada S;2007;J.Phys.Soc.Japan;76;094004
% Derivation of the Quantum Probability Rule without the Frequency Operator
% Journal of the Physical Society of Japan Vol. 76, No. 9, September, 2007, 094004

\bibitem{everett3}
\rfprep\nn Tegmark M;2009;arXiv:0905.2182

\bibitem{cdl}
\rf\nnn Coleman S R, \nn De Luccia F;1980;PRD;21;3305

\bibitem{bubblereview}
\rf\nn Aguirre A,\nn Johnson M; 2009;arXiv;0908;4105A

\bibitem{Vilenkin94}
\rf\nn Vilenkin A;1994;PRL;72;3137


\bibitem{Linde94a}
\rf\nn Linde A;1994;PRD;50;2456

\bibitem{VilenkinVanchurin}
\rf\nn Vanchurin V, \nn Vilenkin A, \nn Winitzki S;2000;PRD;61;083507


\bibitem{Sakaietal00}
\rf\nn Sakai N \nn Nakao K\multiand\nn Harada T;2000;PRD;61;127302

\bibitem{winitz01}
\rf \nn Winitzki S;2002;PRD;65;083506


\bibitem{gratton10}
\rf\nn Gratton S;2010;arXiv;1003;2409

\bibitem{Bellido+98}
\rf\nn Garcia-Bellido J \nn Garriga J\nn Montes X; 1998; PRD;57;4669

%%%


\end{thebibliography}

\end{document}